# Deviations of the distributions of seismic energies from the Gutenberg-Richter law

V.Pisarenko[1], D.Sornette[2,3] and M.Rodkin[1]

[1]International Institute of Earthquake Prediction Theory and
Mathematical Geophysics, Russian Ac. Sci. Warshavskoye sh., 79, kor.2
Moscow 113556, Russia Federation. E-mail: pisarenko@yaol.ru;
rodkin@wdcb.ru

[2]Institute of Geophysics and Planetary Physics
and Department of Earth and Space Science
University of California, Los Angeles, California 90095.
E-mail: sornette@moho.ess.ucla.edu

[3]Laboratoire de Physique de la Matière Condensée
CNRS UMR6622 and Université des Sciences, B.P. 70, Parc Valrose
06108 Nice Cedex 2, France

*Abstract.*

A new non-parametric statistic is introduced for the characterization of deviations of the distribution of seismic energies from the Gutenberg-Richter law. Based on the two first statistical log-moments, it evaluates quantitatively the deviations of the distribution of scalar seismic moments from a power-like (Pareto) law. This statistic is close to zero for the Pareto law *with arbitrary* power index, and deviates from zero for any non-Pareto distribution. A version of this statistic for discrete distribution of quantified magnitudes is also given. A methodology based on this statistics consisting in scanning the lower threshold for earthquake energies provides an explicit visualization of deviations from the Pareto law, surpassing in sensitivity the standard Hill estimator or other known techniques. This new statistical technique has been applied to shallow earthquakes ($h \leq 70$ km) both in subduction zones and in mid-ocean ridge zones (using the Harvard catalog of seismic moments, 1977-2000), and to several regional catalogs of magnitudes (California, Japan, Italy, Greece). We discover evidence for log-periodicity and thus for a discrete hierarchy of scales for low-angle dipping, low-strain subduction zones with a preferred scaling ratio $\gamma=7\pm1$ for seismic moments, compatible with a preferred scaling ratio of 2 for linear rupture sizes, and consistent with previous reports. We propose a possible mechanism in terms of cascades of fault competitions.

*1.**Introduction***

The famous Gutenberg-Richer (G-R) size-frequency law gives the number $N$ of earthquakes of magnitude larger than $m_W$ (in a large given geographic area over a long time interval) (*Gutenberg and Richter*, 1954). Translating the magnitude $m_W=(2/3)(\log_{10} M_W - 16.1)$ in seismic moment $M_W = \mu\, U\, A$ expressed in *dyne-cm* (where $\mu$ is an average shear elastic coefficient of the crust, $U$ is the average slip of the earthquake over a surface $A$ of rupture), the G-R law gives the number $N(M_W)$ of earthquake of seismic moment larger than $M_W$. The striking empirical observation is that $N(M_W)$ can be modeled with a very good approximation by a power law

(1a) $$N(M_W) \sim 1/(M_W)^{\beta},$$

The Gutenberg-Richter law (1a) is found to hold over a large interval of seismic moments ranging from $10^{20} \div 10^{24}$ ($m_W=2.6$-$4$) to about $10^{26.5}$ *dyne-cm* ($m_W=7$). Many works have investigated possible variations of this law (1a) from one seismic region to another and as a function of magnitude and time. Two main deviations have been reported and discussed repeatedly in the literature:

1) from general energy considerations, the power law (1a) has to cross-over at a "corner" magnitude to a faster decaying law. This would translate into a downward bend in the linear frequency-magnitude log-log plot of (1a). The corner magnitude has been estimated to be approximately 7.5 for subduction zones (SZ) and 6.0 for mid-ocean ridge zones (MORZ) (*Pacheco et al.,*1992), (*Okal and Romanowicz,* 1994) but this is hotly debated (see below);

2) the exponent $\beta$ is different in SZ and in MORZ (*Pisarenko and Sornette,* 2003a). There is in addition a controversy among seismologists about the homogeneity of $\beta$-values in different zones of the same tectonic type. Some seismologists believe that $\beta$-values are different at least in several zone groups, others find these differences statistically insignificant.

The authors ( *Main and Burton*, 1984; *Rundle*, 1989; *Romanowicz*, 1994; *Pacheco et al.*, 1992; *Pacheco and Sykes*, 1992; *Romanowicz and Rundle*, 1993; *Okal and Romanowicz*, 1994; *Sornette et al.*, 1996; *Kagan*, 1997; 1999; *Sornette and Sornette*, 1999) proposed that the large-magnitude branch of the distribution can be modeled also by a power-like law and that the crossover moment or magnitude between these two distributions can be connected with the thickness of the seismogenic zone. *Pacheco et al.* (1992) claimed to have identified a kink in the distribution of shallow transform fault earthquakes in MORZ around magnitude 5.9 to 6.0, which corresponds to a characteristic dimension of about 10 km; a kink for subduction zones is presumed to occur at a moment magnitude near 7.5, which corresponds to a downdip dimension of the order of 60 km. However, *Sornette et al.* (1996) have shown that this claim cannot be defended convincingly because the crossover magnitude between the two regimes is ill-defined. *Pisarenko and Sornette* (2003a,b) suggested new statistical tests to find deviations of earthquake energy distribution from the Gutenberg-Richter law at extreme range and use the Generalized Pareto Distribution (GPD) to characterize tails of energy distribution in this range. In particular, using a transformation of the ordered sample of seismic moments into a series with uniform distribution under the assumption of no crossover and applying the bootstrap method, *Pisarenko and Sornette* (2003b) estimated a crossover magnitude $m_W$=8.1± 0.3 for the 14 subduction zones of the Circum Pacific Seismic Belt. Such a large value of the crossover magnitude makes it difficult to associate it directly with a seismogenic thickness as proposed by many different authors in the past. The present paper can be seen as a continuation of the study begun in (*Pisarenko and Sornette*, 2003a,b), but we stress that this continuation is based on a quite different non-parametric approach, in contrast to the parametric methods previously developed.

There is a second important novelty here. Complementary to these previous works emphasizing deviations from the G-R distribution only in the tail, the present paper explores the possible existence of deviations from (1a) elsewhere, that is, in the bulk of the distribution. This becomes possible by our introduction of a new statistic, which turns out to be very sensitive to deviations from a pure power law. Namely, we suggest a non-parametric statistic that is close to zero for "pure" G-R laws, and deviates from zero at energy sub-ranges with large enough deviations from the G-R law. This is a first attempt to address the problem of characterizing deviations in the whole range of seismic moment sizes, including moderate and small events, with the hope of connecting such hypothetical deviations with tectonic and geological particularities of the zones in question.

Deviations of a pure G-R power law can take a priori many shapes. It has been recognized with the development of the concept of fractals (*Mandelbrot*, 1982) that power laws are the hallmark of the symmetry of continuous scale invariance (CSI) (*Dubrulle et al.*, 1997 and references therein). Deviations of a pure G-R power law thus express some degree of breaking of this CSI symmetry. As for any other symmetry, there are many ways to break the CSI symmetry. One of them is particularly interesting because it constitutes a minimalist way of breaking the CSI symmetry: it corresponds to keeping the scale invariance but only for specific scales organized according to a discrete hierarchy with some fixed preferred scaling ratio γ. The lower symmetry thus obtained is called "discrete scale invariance" (DSI) (*Sornette*, 1998 and references therein). Going from CSI to DSI corresponds to a partial breaking of the CSI, conceptually similar to the partial breaking of continuous translational invariance in liquids into discrete translational invariance in solids. In the present paper, our new statistic unearths a DSI structure decorating the G-R power law, which is the most apparent for low-strain low-angle dipping subduction zones. This is particularly interesting because it complements from a novel angle with a different data set previous reports of DSI in crack growth (*Ouillon et al.*, 1996; *Huang et al.*, 1997), rupture and fragmentation (*Sadovskii*, 1999; *Geilikman and Pisarenko*, 1989; *Sahimi and Arbabi*, 1996; *Johansen and Sornette*, 1998; *Suteanu et al.*, 2000); and seismicity (*Sornette and Sammis*, 1995; *Newman et al.*, 1995; *Saleur et al.*, 1996). We elaborate on the implication of this finding in the discussion section.

The organization of this paper is as follows. In section 2, we describe the new statistic tailored for studying deviations from the G-R law and summarize its main properties (for continuous variables such as seismic moment). In Section 3, we present a similar technique for catalogs with quantized magnitudes, as they are usually given in seismic catalogs (for instance in 0.1 magnitude units). In Section 4, we apply these statistics both to some simulation examples and to the Harvard catalog of seismic moments. In Section 5, we apply the discrete version of our statistic to several regional catalogs. Section 6 presents a discussion of our results and concludes.

## 2. TP-statistic and its properties.

It is well-known that, in terms of (scalar) seismic moments, the G-R law coincides with the Pareto distribution $F(x)$, allowing us to rewrite (1a) as

(1) $\qquad F(x) = 1 - (u/x)^\beta, \; x \geq u, \; \beta > 0,$

where $u$ is a lower threshold, and $\beta$ is the power index of the distribution. Let us consider a finite sample $x_1, \ldots, x_n$. It is desirable to construct a statistic $TP = TP(x_1, \ldots, x_n)$ such that, asymptotically for large $n$, $TP$ would be close to zero and, at the same time, would deviate from zero for samples whose distribution deviates from eq.(1). Let us construct such statistic based on the first two normalized statistical log-moments of the distribution (1). Using the symbol E for the mathematical expectation, we have

(2) $\quad E \log(X/u) = \int_u^\infty \log(x/u) \, dF(x) = 1/\beta;$

(3) $\quad E \log^2(X/u) = \int_u^\infty \log^2(x/u) \, dF(x) = 2/\beta^2.$

Thus, if we choose

(4) $\qquad TP = (1/n) \sum_{k=1}^n \log(x_k/u))^2 - (0.5/n) \sum_{k=1}^n \log^2(x_k/u),$

then according to the Law of Large Numbers and equations (2)-(3), the statistic $TP$ tends to zero as $n \to \infty$. In order to evaluate the standard deviation $std(TP)$ of the statistic $TP$, we rewrite (4) in the form:

(5) $\qquad TP = (1/n) \sum_{k=1}^n [\log(x_k/u) - E_1] + E_1)^2 - (0.5/n) \sum_{k=1}^n [\log^2(x_k/u) - E_2] - 0.5 E_2,$

where $E_1, E_2$ are the expectations of $\log(x_k/u)$ and $\log^2(x_k/u)$ respectively (for Pareto samples, $E_1 = 1/\beta$ and $E_2 = 2/\beta^2$). Both sums in eq.(5) are of the order $n^{-0.5}$:

$$\varepsilon_1 = 1/n \sum_{k=1}^n [\log(x_k/u) - E_1] \propto n^{-0.5}; \; \varepsilon_2 = 1/n \sum_{k=1}^n [\log^2(x_k/u) - 2/\beta^2] \propto n^{-0.5}$$

Thus, if $n$ is large enough, we can expand *TP* in eq.(5) into Taylor series up to terms of the order $n^{-0.5}$ in the neighborhood of $E_1$ and $E_2$ respectively:

(6) $$TP \cong (E^2_1 - 0.5 E_2) + 2 E_1 \varepsilon_1 - 0.5\, \varepsilon_2 .$$

This provides an estimation of *std(TP)* by the standard deviation of the sum:

(7) $$2 E_1 \varepsilon_1 - 0.5\, \varepsilon_2 = (2 E_1 / n) \sum_{k=1}^{n} [\log(x_k/u) - E_1] - (0.5/n) \sum_{k=1}^{n} [\log^2(x_k/u) - E_2] =$$

$$= (0.5 E_2 - 2 E^2_1) + (1/n) \sum_{k=1}^{n} [2 E_1 \log(x_k/u) - 0.5 \log^2(x_k/u)].$$

The standard deviation of the last sum in (7) can be estimated by

(8) $$n^{-0.5}\, std[2 E_1 \log(x_k/u) - 0.5 \log^2(x_k/u)],$$

and the standard deviation *std* of the term in bracket in eq.(8) is estimated through its sampled value *[2 $E_1$ log($x_k$/u) – 0.5 log²($x_k$/u)]*. Equation (8) provides an estimate of *std(TP)* if we replace $E_1$ by its sample analog:

$$(1/n) \sum_{k=1}^{n} \log(x_k/u).$$

Fig. 1 shows the *TP*-statistic as function of the lower threshold *u,* applied to a simulated Pareto sample of size *n=5000* with power index $\beta = 2/3$ generated with *u*=1 as defined in Eq.(1). Keeping fixed the synthetically generated data, for a given lower threshold *u*, we select all data values that are larger than *u* and calculate *TP* using only these values above *u*. Varying *u* allows in principle to test different part of the distribution. If the lower threshold *u* is smaller than *300*, the *TP*-statistic does not deviate significantly from zero. Beyond this value (for which there are less than 150 data values), random fluctuations become large as *TP* is estimated with a smaller and smaller number of data.

Fig. 2 shows the *TP*-statistic applied to a sample generated with the distribution function *F(x)* formed by two jointed Pareto branches with different power indices:

(9) $$F(x) = \begin{cases} 1 - 1/x^{b_1}; & 1 \leq x \leq c; \\ 1 - c^{b_2-b_1}/x^{b_2}; & c \leq x, \end{cases}$$

with $\beta_1 = 2/3$, $\beta_2 = 3.5$ and c=300. The behavior of the *TP*-statistic in this case is quite different from that observed for a single pure power shown in Fig. 1. First of all, in the lower and middle part of the range, *TP* is shifted up by *0.2 – 0.3*. Then, beyond the value *h*

= *100*, a rather steep descent starts ending with almost zero value at *h=300*. The difference between these two distributions thus leads to strong distinguishing signatures.

Having in mind the problem of DSI that we shall encounter in our investigations below, we illustrate the application of the *TP*-statistic to simulated samples generated by a Pareto-like distribution with log-periodic oscillating deviations from an exact Pareto law. Namely, the chosen distribution function is defined by:

(10) $$F(x) = \begin{cases} 1 - C_1(b,\Delta b)/x^{b+\Delta b}; & k \cdot \Delta l \leq \log_{10}(x) < (k+1/2) \cdot \Delta l; \\ & k=0,1,... \\ 1 - C_2(b,\Delta b)/x^{b-\Delta b}; & (k+1/2) \cdot \Delta l \leq \log_{10}(x) < (k+1) \cdot \Delta l. \end{cases}$$

The theoretical tail of this distribution function with *b=0.67; Δb=0.2; Δl=0.75* is shown in Fig.3a (*Δl* is the $\log_{10}$-period of the DSI oscillations, as it is seen from Eq.(10), corresponding to a preferred scaling ratio $10^{\Delta l}=5.62$); $C_1$, and $C_2$ are normalizing constants depending on *b,Δb*. With *Δb=0.2,* the oscillations are hardly observable on the graph. Fig.3b shows a sample analog of the tail function constructed from a simple of size *n=5000* generated with the DF (10). Fig.3c shows the *TP*-statistic as a function of the lower threshold *u* applied to the above sample. We see that the DSI oscillations are very strong and distinguishable despite some noise disturbances (in particular, at large lower thresholds *u*). Note that the maxima and minima of the log-periodic oscillations shown in Fig.3c correspond to the points of changing slopes shown in Fig3a: local maxima of Fig.3c correspond approximately to the transition from slope *(b+Δb)* to slope *(b-Δb),* and local minima to the transition from *(b-Δb)* to *(b+Δb)*.

### *3. TED-statistic for discrete exponential distribution.*

Many catalogs measure earthquake sizes with discrete magnitudes rather than with continuous seismic moments. Accordingly, the G-R law is an exponential distribution when earthquake sizes are expressed in magnitudes. In this section, we address the problem of constructing a statistic similar to *TP* for a quantized exponential distribution. Let us consider the (shifted) exponential distribution:

(11) $$F(x) = 1 - \exp(-(x-u)/d); \quad x \geq u,$$

where *d* is a scale parameter. Let us assume furthermore that the distribution (11) is quantified with a step *Δ* providing discrete probabilities:

(12) $$p_k = F(u+k\Delta) - F(u+(k-1)\Delta) = \exp(-(k-1)\Delta/d) - \exp(-k\Delta/d), \quad k = 1,2,... \ .$$

Suppose further that, in a given catalog of magnitudes, there are $m_k$, $m_k \geq 0$, values within the interval $(u+(k-1)\Delta\,;\,u+k\Delta)$, $k=1,2,...$, so that

(13) $\quad m_1 + m_2 +\ldots+m_k +\ldots = n,$

where $n$ is the total number of observed magnitudes (sample size). We can consider the sample $(m_1, m_2, \ldots m_k \ldots)$ as the result of $n$ independent trials of discrete rv $\xi$ taking values $1,2,...,k,...$ with probabilities (12). The first two moments of the rv $\xi$ are:

(14) $\quad M_1 = \sum_{k=1}^{\infty} k\, p_k = 1/(1-\exp(-\Delta/d))\,; \quad M_2 = \sum_{k=1}^{\infty} k^2\, p_k = (1+\exp(-\Delta/d))/(1-\exp(-\Delta/d))^2.$

Denoting $\lambda = \exp(\Delta/d)$ we get:

(15) $\quad M_1 = \lambda/(\lambda-1); \quad M_2 = \lambda(1+\lambda)/(\lambda-1)^2.$

We derive from the first equation in (15):

(16) $\quad \lambda = M_1/(M_1 - 1),$

and from the second equation in (15):

(17) $\quad \lambda = (M_1 + M_2)/(M_2 - M_1).$

Replacing the theoretical moments $M_1$, $M_2$ by their sample analogs

(18) $\quad M^*_1 = \sum_{k=1}^{\infty} k\, m_k/n\,; \quad M^*_2 = \sum_{k=1}^{\infty} k^2\, m_k/n,$

in Eqs.(16),(17), we get two different estimates of $\lambda$ whose difference converges to zero in probability as $n \to \infty$. This results from the fact that $M^*_1, M^*_2$ are consistent estimates of $M_1, M_2$ for any value of the unknown scale parameter d. This provides the looked-for *TED*- statistic:

(19) $\quad TED = (M^*_1 + M^*_2)/(M^*_2 - M^*_1) - M^*_1/(M^*_1 - 1).$

Our remaining task is to derive a consistent sample estimate of the standard deviation *std(TED)* of the statistic *TED* defined by (19). For this purpose, we use the formulae derived in *(Rao,* 1966*)* for the variance of limit normal distributions for sample moments of a multinomial distribution. The variance of the limit normal distribution of *TED* can be estimated as follows. Let us denote by $U_1$, $U_2$ the following statistics*:*

(20)  $U_1 = 1/(M^*_1 - 1)^2 + 2/(M^*_2 - M^*_1) + 2M^*_1/(M^*_2 - M^*_1)^2$ ;

(21)  $U_2 = 2M^*_1/(M^*_2 - M^*_1)^2$ .

Then, the variance of the limit normal distribution of *TED* can be consistently estimated by the following expression (see for details *(Rao, 1966)*, Chapter 6, Section 6a.1):

(22)  $var(TED) \cong \sum_{k=1}^{\infty} k^2 (m_k/n)(U_1 - kU_2)^2 - [\sum_{k=1}^{\infty} k(m_k/n)(U_1 - kU_2)]^2$ .

## 4. Application of TP-statistic to the Harvard catalog of seismic moments.

Let us consider first the Harvard catalog of (scalar) seismic moments for the time period 1977-2000, and for shallow events ($h \leq 70$ km). Only earthquakes with seismic moments larger than $10^{24}$ *dyne-cm* are considered to ensure a tolerable completeness and homogeneity. The total number of earthquakes in all 14 considered subduction zones (SZ) is 4609, whereas the total number of events in midocean ridge zones (MORZ) is 1286. More detailed information on the seismic moment data can be found in (*Pisarenko and Sornette,* 2003a).

### 4.1. TP-statistic of earthquakes in subduction zones (SZ)

The sample tail of SZ events is shown on Fig. 4a. Except for the extreme range, the sample tail function *1-F(x)* looks like a straight line in double log-scale, i.e. the G-R law seem to apply. At the very extreme end of the range of seismic moments, a "bend down" can be observed, but there is no strikingly visible deviations from a straight line in the middle part (the careful reader may however notice the existence of an oscillation of very small amplitude). Fig. 4b shows the *TP*-statistic applied to the subduction sample. Regular oscillations on a noisy background are now obvious. In addition, the TP-statistic is translated upward by 0.1-0.2 which, by comparison with the synthetic test (9) shown in Fig.2, is probably the signature of the bend down in the extreme range. By comparing Fig. 4b with Fig. 1, it is clear that the *TP*-statistic with a moving lower threshold provides a rather sensitive method for detecting deviations from the G-R law. We can already conclude from this analysis that the distribution of earthquake moments exhibits significant deviation from the pure power law, not only in its tail but also, in a major portion of the scaling region. In a quantitative form, this evidence can be considered as a new claim. The comparison of the oscillations observed in Fig. 4b with those of the synthetic test in Fig. 3c suggests that the deviation of the distribution of seismic moments from the G-R law can be modeled by log-periodic oscillations, reflecting a DSI partial breaking of CSI. For comparison, we show on Fig. 4c the so-called Hill's estimates of the power index $\beta$ as function of the moving lower threshold $u$ (see e.g. *Embrechts et al.,*1997, for details on the Hill's estimators). The oscillations of the power index are hardly observable on this graph.

### *4.2. TP-statistic of earthquakes in mid-ocean ridge zones* (MORZ)

Earthquakes in MORZ can be divided into two main groups according to their source mechanism: events with normal fault mechanism (spreading areas), and events with strike-slip mechanism (transform areas). These two groups differ quite remarkably by their source mechanism as well as by their total energy. We thus treat them separately. We have performed the corresponding classification using the technique of *triangle diagram*, as described in (*Kaverina et al.,*1996). A small number of events could not be classified with this technique due to uncertain characteristics of the source. We thus obtain 926 transform events and 360 spreading events. The corresponding sample tail functions are shown in Figs. 5 and 6. In these cases, deviations from the G-R law can be seen with the naked eye: the tail functions are convex, except for the four largest spreading events in Fig. 6 that should be discussed separately (this goes beyond of scope of the present paper). Figs. 7 and 8 show the *TP*-statistic for these two MORZ samples. As it could be expected, deviations from zero are significant in both cases, confirming significant deviations from a pure power law both in the bulk and in the tail of the distributions.

### *4.3. DSI in the TP-statistic of earthquakes in subduction zones*

Let us come back to the oscillations observed in Fig. 4b for the distribution of event sizes in subduction zones. While they are significant, can they be related to any geophysical characteristics? We are going to suggest such a tentative geophysical interpretation if these oscillations are genuine, acknowledging in the same time that a definite conclusion on this subject needs more detailed study based on a more representative data.

The seismic and stress-strain regimes in active transitional zones are believed to be determined mainly by the mechanical coupling between the downgoing slab and the overriding continental plate. The dip angle is known as one of factors affecting the mechanical coupling (*Jarrard*, 1986). It seems reasonable to suggest that the increase in the mechanical coupling promotes both an increasing seismic activity and a more noticeable log-periodicity.

To perform a quantitative analysis, we use a pre-existing classification of subduction zones constructed by Jarrard *(Jarrard*, 1986), based on a set of geological and tectonic characteristics (29 parameters), of which we select the following main 5 parameters:

-intermediate dip up to 100 km of depth (in degrees);

-strain class (in an abstract discrete scale from 1 to 7);

-convergence rate (in cm/year);

-mean slab age at trench (in m.y.);

-maximum moment magnitude $M_w$.

Jarrard (1986) provided the corresponding 5 parameter values for each of his 39 subduction zones. To obtain the 5 parameter values for each of our 14 subduction zones, each one often made of several zones considered by Jarrard, we averaged the corresponding parameter values over all Jarrard's zones constituting each of our zones. The resulting values of the 5 parameters for each of our 14 subduction zones 5 are given in Table I. We then consider each of the 5 parameters one by one. For each parameter, we classify all 14 zones into two groups with approximately equal numbers of elements in accordance with "relatively high" (H), or "relatively low" (L) values of the parameter. This provides us with 5 different partitions of 14 zones marked in Table I as (L) and (H). We group all earthquakes of the zones in the "high" (respectively "low") group of a given partition and apply the $TP$-statistic to this "high" (respectively "low") group separately. Fig. 9a shows that the group with "low" dips is characterized by significant log-periodic oscillations of its $TP$-statistic, whereas no oscillation can be observed in the $TP$-statistic of the "high" dip group (Fig. 9b). Similar differences in the $TP$ statistics of the "low" and "high" strain classes are found (see Fig. 10a and 10b). The classification using the three other parameters does not give noticeable differences between the H and L classes. This suggests that the log-periodic oscillations are associated with low dip and low strain subduction zones in the middle part of the slab (60-100 km depth). From the measure of the period in the logarithm of the lower threshold, we infer a preferred scaling ratio $\gamma=7\pm1$. In order to compare the sensitivity of the $TP$-statistic and of Hill's estimates, Fig. 11a and Fig. 11b show Hill's estimates of the power index $\beta$ for low and high dip, that should be compared with Figs. 9a and 9b. Oscillations are hardly seen, if any.

The origin for log-periodicity may be due to a mechanism similar to that found in growing antiplane shear faults (*Huang et al.*, 1997) according to the following mechanism. With simple subduction zone plate bending, intraplate outer rise earthquakes are mostly due to tensional failure at shallow depths. The location of the plate bending corresponds to the location of stress concentration constituting the most favorable loci for earthquake nucleation and fault growth. We visualize a system of faults more or less parallel to the subduction boundary. As the plate undergoes its subduction, faults compete with each other to accommodate the growing strain: nearby faults screen each other. This competition between two neighboring faults imply that one of them will start to grow faster and be more active while the other one slows down, being screened by the first one. As this process can occur at all scales, this leads to a cascade of Mullins-Sekerka instability as demonstrated in (*Huang et al.*, 1997) by analytical as well as numerical calculations: from an initial homogenous population of faults, the cascade of growth instabilities create a discrete hierarchy of fault lengths with a scaling factor between successive levels of the hierarchy close to 2. Thus, we can assume that earthquakes reveal these discrete hierarchy of faults. As seismic moments are approximately proportional to the cube of the rupture length, this predicts a preferred scaling factor of $2^3=8$ for seismic moments. This value is compatible with our measurement $\gamma=7\pm1$, as seen from Figs. 9a and 10a. We stress that the mechanism of competition between growing and reactivated faults operates both for normal as well as for transform faults and is not restricted to antiplane shear faults. Our proposed mechanism rationalizes our finding that a DSI fault network should be more apparent for low-strain low-angle dipping subduction: only then can a large delocalized lateral zone of faulting be created with many sub-parallel faults interacting and competing. Large dipping angles localize the region of competing faults to a narrow scale range, preventing the observation of log-periodicity.

We recall that all these considerations are tentative since the number of different subduction zones under analysis (14) is too small to draw a definite conclusion.

## 5. *Application of TED-statistic to catalogs with discrete magnitudes.*

Let us now illustrate the application of the *TED*-statistic introduced in section 3 to the global NEIC catalog as well as to the regional catalogs of Japan, California, Italy and Greece, all reported with discrete magnitudes (the discrete magnitude bins are in all cases *0.1* of the magnitude unit). The sample magnitude-frequency curves (MFC) and corresponding TED-statistic as a function of the lower threshold $u$ are shown on Figs. 12a-e and 13a-e correspondingly.

On the MFC of the global NEIC catalog, there is no visible feature except for a downward bend at the extreme range, starting near *M=7.5*. On the *TED*-graph, there is a corresponding increasing *TED*-statistic with an acceleration at the extreme range. Thus, the tail of the sample MFC deviates more and more strongly downward from the G-R law as one penetrates further in the tail of the distribution. One could say that this global catalog is "too heavily averaged" to find any other significant characteristic deviation from the G-R in the middle part of range.

Regional catalogs reveal more structures in their *TED*-graphs. The sample MFC for Japan is visibly a convex function with a small downward bend at the extreme range. On its corresponding *TED*-curve, besides the extreme upward bend, one sees two small "troughs" near $M \cong 5.75$ and $M \cong 6.9$. The nature of these troughs is unclear. In the small and intermediate range *3 ≤ M ≤ 5.75*, the *TED*-curve is almost horizontal (a small negative shift is probably due to the final downward bend of the MFC). It thus seems that the G-R law for Japan in the middle part of range is fulfilled satisfactorily.

The MFC of the California catalog has a slightly convex form in the range *1 ≤ M ≤ 3* that can be explained by the partial incompleteness of the catalog in this range. This fact is reflected by a positive deviation of the *TED*-curve in this interval. Above $M \cong 3$, the completeness of the catalog seems satisfactory, and the *TED*-curve remains near zero till $M \cong 5$, where a smooth increase starts interrupted by two "humps" near $M \cong 5.9; 6.3$ (a "trough" near $M \cong 6.7$ does not seem reliable because of the large *std*). In the central part of the range *3 ≤ M ≤ 5*, deviations from the G-R law are small.

The MFC of Italy exhibits some deviations caused by the practice of a human operator to prefer magnitude values that are multiples of *0.5*. Small local minima corresponding to such magnitude values are hardly detectable in the MFC, but they are clearly observed in the *TED*-curve, sometimes with small shifts. Indeed, one can discern small but quite definite minima near $M \cong 3; 3.5; 4.1; 4.6; 5; ....$ A "hump" near $M \cong 5.7\text{-}5.8$ is seen, but it cannot be ascertain due to the large *std*.

The MFC of Greece exhibits the same "operator" minima as described for Italy at $M \cong$ *4.1; 5.1; 6.1*. They are "compensated" by neighboring maxima. It seems evidently, that these minima/maxima are not connected with natural sources, but what is strange is that they are shifted by *≈0.1-0.2* magnitude unit with respect to integer numbers. The reason of this shift is not quite clear. One possible explanation can be following. The regional seismological networks rather frequently use the local $M_L$ magnitude scale and $m_b$ values obtained from the high-frequency seismic wave range. For the examined magnitude range (from *4* up to *5.5*), if being recalculated to the world-wide used low-frequency $m_b$ scale both $M_L$ and high-frequency $m_b$ values will be increased by about *0.2*, see for example, (*Sobolev*, 1993). The *TED*-curve behaves quite satisfactorily except for the small minima/maxima mentioned above and the steep upward bend at the extreme range. With this exception, deviations from the G-R law in the range *3.4 ≤ M ≤ 6.7* are not large.

Thus, unlike the Harvard catalog of seismic moments, we did not find noticeable oscillations in magnitude catalogs. The four local catalogs (Japan, California, Italy and Greece) are characterized by very complex boundaries and we would not expect the simple mechanism of competing faults to be as clearly *apparent* due to the probable effect of averaging over different fault orientation and fault mechanism: indeed, previous studies have shown that log-periodicity can only reveal DSI if adequate steps to prevent the destruction of the oscillations by averaging are taken (*Johansen and Sornette*, 1998). The situation is not improved by the use of discrete magnitudes as it is well known that techniques of measuring earthquake size by magnitude is less accurate as compared with the method of seismic moments. Different Magnitudes are based on measurements of maximum amplitudes on seismograms registered by seismometers with different frequency ranges. Besides, many magnitude catalogs have had a long evolution of measurement procedures, thus causing some non-stationarity in the registered time-series of observed magnitudes. Quantifying magnitudes with step *0.1* can lead to inaccuracies for the determination of earthquake energies, but probably with not too serious consequences (except for the measurement of log-periodicity). Of course, larger steps, say of *0.25*, are able to cause appreciable undesirable effects. To check this, we have converted the Harvard seismic moments (for all 14 subduction zones, *n=4609*) into discrete magnitudes in units of 0.1 and have applied to them our TED-statistic (see Fig. 14). We obtain very weak maxima which, while corresponding to the strong maxima described above for the continuous TP-statistic, are barely statistically significant with the TED-statistic. We observe the same phenomenon with synthetic log-periodic power laws of seismic moments, when transformed in discrete magnitudes. Thus, the quantization of magnitudes decreases strongly the efficiency of the detection of log-periodic oscillations. Thus, for detailed statistical analysis of deviations from the G-R law, the Harvard catalog of seismic moments is preferable to the magnitude catalogs, despite the fact that magnitude catalogs cover longer time intervals.

In order to complete comparison of discrete magnitudes with continuous seismic moments on the same events, we have taken magnitude values reported in the Harvard catalog just for the same events that were used in our subduction sample (*n=4609*). There are two magnitudes in the Harvard catalog: magnitude $m_b$ determined from body wave measurements, and magnitude $M_s$ determined from surface waves. Unfortunately, $M_s$ is reported only for 88-90% of all events, which makes unreasonable using this magnitude for our comparison. Thus, we compared the seismic moments *M* with two versions of magnitudes: $m_b$ and $M_{max} = max(m_b; M_s)$. Fig. 15a,b show sample tails of distributions of these magnitudes. We see that the $m_b$-tail is a convex curve, thus visibly disobeying the G-R law, whereas the $M_{max}$-tail looks in the middle range *(5.5; 7.5)* rather satisfactorily. Fig. 16a,b show the *TED*-statistics for these two magnitude distributions. On the *TED* for $m_b$, weak traces of peaks at $m_b \cong 6.1; 6.6; 7.1$ still can be discerned, whereas there is no visible peaks at these magnitudes on the *TED* for $M_{max}$. Thus, we can conclude once more that the catalog of seismic moments is preferable for detailed statistical treatments as compared with magnitude catalogs.

## 6. Discussion and Conclusions

We have suggested a *T*-statistic of a new type measuring quantitatively the deviations from the G-R law expressed both as a function of seismic moments and as a function of discrete magnitudes. This statistic can be displayed as a function of a lower threshold. In this respect, it is similar to the well known *mean excess function e(u),* see e.g. (*Embrechts et al.,*1997):

(23)  $e(u) = E( X-u \mid X > u)$.

This function was introduced as a useful statistical tool to characterize tails of distributions. For power-like tails, it behaves as a straight line with positive slope; for exponential tails, it is a constant, and so on. Our statistic (for continuous distributions) is based on two *mean log-excess moments* $l_1(u), l_2(u)$:

(24)  $l_1(u) = E ( log(X/u) \mid X > u); \quad l_2(u) = E ( log^2(X/u) \mid X > u)$.

For discrete magnitudes with exponential distribution (the G-R law), in addition to (23), the second *mean excess moment* $e_2(u)$ in discrete form was used:

(25)  $e_2(u) = E( (X-u)^2 \mid X > u)$.

These *T*-statistic based on statistical moments are not local characteristics of corresponding densities, they are characteristics of a *cumulative* type, referring to the tail on the interval $(u, \infty)$. Thus, they do not answer to question "at what location a particular sample differs from the G-R law?", but rather to "what part of the extreme tail differs from the G-R law?". We gave examples of the application of the *T*-statistics to several previously well-studied earthquake catalogs. Because of its cumulative property, the *T*-statistic is, generally speaking, more stable than local characteristics of deviations from a given law. The *T*-statistic permits to judge on the deviation of an integral tail portion from the G-R law, but it is not tailored specially to work with the limit extreme behavior of the tail. For the latter, we had suggested other methods in *(Pisarenko and Sornette,* 2003a,b), and we hope to continue working in this direction elsewhere.
    Application of the *TP*-statistic to subduction zones (the Harvard catalog of seismic moments) permitted to discover some heterogeneities in the seismic moment-frequency curve. They are exhibited more explicitly for subduction zones with low dip angle and with low stress. We cannot say definitely what is the nature of these heterogeneities which appear as oscillations of the *TP*-statistic in the logarithm of the lower threshold. We can't say that we are absolutely free from some doubts that this discovered oscillations are artifact due to some peculiarities of the algorithms used in the data processing of the

Harvard catalog. We can't find neither any definite reason for considering our discovered oscillations as an artifact. So, with these reservations, we shall consider them as a natural effect. As an explanation of these log-periodic oscillations we can accept that they are the signature of a discrete scale invariance (DSI) in the seismic catalogs. We suggested a simple mechanism in terms of competing faults localized in the domain where the bending of the subducting plate is concentrated.

More generally, if one agrees that the observed log-periodicity is not an artifact and has a natural cause, we would like to emphasize that log-periodicity is an inherent property of systems *with discrete self-similarity,* see *(Geilikman and Pisarenko,* 1989; *Sornette and Sammis,* 1995; *Sornette,* 1998, *Sornette,*2000*)*. Thus, we can argue that some physical field(s) with discrete self-similarity underlies these oscillation effects. These ideas are close to those suggested by M.Sadovskii in explaining numerous examples of "preferable sizes" in nature: geological blocks, ground particles, dimensions of rock pieces produced by explosions, celestial bodies etc., see *(Sadovskii,*1999*).* In all these examples, one observes some "humps" of preferred sizes on a smooth background of the distributions. According to Sadovskii, the mean log-distance between two neighboring humps (for linear dimensions of objects, or reduced to linear dimensions) varies from *$log_{10} 2 = 0.3$* to *$log_{10} 4.5 = 0.65$* with average value *$log_{10} 3 \cong 0.5$.* If we take our "humps" on the *TP*-curves for subduction zones and transform their values into linear dimensions (assuming that earthquake energy is proportional to the cubic linear dimension of the source), we get a log-distance about *0.33* (preferred scaling ratio close to 2 for length scales) which falls into the interval indicated by Sadovskii.

Catalogs with discrete magnitudes (quantified with *0.1* of magnitude units) did not show noticeable oscillations in the *TED*-statistic. A possible reason lies in less precise measurements of earthquake sizes by magnitudes as compared with seismic moments, and in the non-stationarity of catalog time-series. Thus, in our opinion, for detailed statistical analysis of the type performed here, the Harvard catalog of seismic moments is preferable. On the whole, the agreement of regional catalogs of large sizes, say, $10^4$ or more, with the G-R law is satisfactory in all ranges, except at the extreme end which needs a special study.

We leave for the future the use of our *T*-statistic for studies of zonal seismic particularities. This can only be performed under the condition that the corresponding zonal catalog is sufficiently large (perhaps, no less than $10^3$ events, depending on the range in question). For zonal catalogs, deviations from the G-R low (if one believes in zonal validity of the G-R law, and majority of seismologists do believe in such validity) can be connected with tectonic and geological particularities of the zone in question (see for instance (*Bird et al.*, 2002)), which is impossible for global catalogs, or catalogs including several tectonically different areas.

*Acknowledgements.*

This work was supported by Russian Fondation of Basic Research, grant 02-05-64379 (Pisarenko, Rodkin). This work is partially supported by NSF-EAR02-30429, by the Southern California Earthquake Center (SCEC) and by the James S. Mc Donnell Foundation 21st century scientist award/studying complex system.  SCEC is funded by NSF Cooperative Agreement EAR-0106924 and USGS Cooperative Agreement

02HQAG0008. The SCEC contribution number for this paper is 751. We are grateful to A.Lander for valuable discussion and for help in preparing the earthquake catalogs.

FIGURE CAPTIONS

Fig. 1: *TP*-statistic as a function of the lower threshold *u*, applied to a simulated Pareto sample of size *n=5000* with power index $\beta = 2/3$ generated with *u=1* as defined in Eq.(1). Increasing *u* decreases the number of data values used in the calculation of the statistic *TP*, thus enhancing the fluctuations around 0. The two dashed lines show plus or minus one standard deviation std estimated as exposed in the text.

Fig. 2: Same as Fig. 1 for a synthetic sample generated with the distribution function *F(x)* formed by two jointed Pareto branches with different power indices defined in Eq.(9) with $\beta_1= 2/3$, $\beta_2= 3.5$ and c=300. In contrast to Fig. 1, *TP* is significantly different from 0 over the whole range, signaling a significant deviation from a pure G-R power law.

Fig.3a : Theoretical tail of the distribution function (10) constructed to have regular log-periodic oscillating deviations around an average pure power law. The parameters are *b=0.67; Δb=0.2; Δl=0.86*. (*Δl* is the $\log_{10}$-period of the DSI oscillations, as it is seen from eq.(10), corresponding to a preferred scaling ratio $10^{\Delta l}=5.62$); $C_1$, and $C_2$ are normalizing constants depending on *b,Δb*.

Fig.3b: Tail of the distribution obtained from a sample generated by the distribution function shown on Fig. 3a. Sample size *n=5000*.

Fig.3c: *TP*-statistic as a function of the lower threshold *u* applied to the sample of size *n=5000* generated with the DF (10) shown in Fig. 3a, with *b=0.67; Δb=0.2; Δl=0.86*.

Fig. 4a: Sample tail of the distribution of seismic moments for subduction zones (SZ) earthquakes from the Harvard catalog of (scalar) seismic moments for the time period 1977-2000, and for shallow events (*h ≤ 70 km)*. Only earthquakes with seismic moments larger than $10^{24}$ *dyne-cm* are considered to ensure completeness and homogeneity. Sample size *n=4609*.

Fig. 4b: *TP*-statistic applied to the subduction sample whose distribution is shown in Fig. 4a.

Fig. 4c: Hill's estimates of the power index $\beta$ as function of the moving lower threshold *u* applied to the subduction sample whose distribution is shown in Fig. 4a.

Fig.5: Sample tail distribution function for the 926 transform events in MORZ classifed using the technique of *triangle diagram*, as described in (*Kaverina et al.,*1996).

Fig.6: Sample tail distribution function for the 360 spreading events in MORZ classifed using the technique of *triangle diagram*, as described in (*Kaverina et al.,*1996).

Fig.7: *TP*-statistic as a function of the moving lower threshold $u$ for the 926 transform events in MORZ whose distribution is shown in Fig, 5.

Fig.8: *TP*-statistic as a function of the moving lower threshold $u$ for the 360 spreading events in MORZ whose distribution is shown in Fig, 6.

Fig.9a: *TP*-statistic as a function of the moving lower threshold $u$ for the group of subduction zones with "low" dip angles.

Fig. 9b: *TP*-statistic as a function of the moving lower threshold $u$ for the group of subduction zones with "high" dip angles.

Fig. 10a: *TP*-statistic as a function of the moving lower threshold $u$ for the group of subduction zones with "low" strain.

Fig. 10b: *TP*-statistic as a function of the moving lower threshold $u$ for the group of subduction zones with "high" strain.

Fig. 11a: Hill's estimates of the power index $\beta$ as a function of the moving lower threshold $u$ for the group of subduction zones with "low" dip angles.

Fig. 11b: Hill's estimates of the power index $\beta$ as a function of the moving lower threshold $u$ for the group of subduction zones with "high" dip angles.

Fig. 12: Empirical tail distributions of earthquake magnitudes: *(a)* global NEIC catalog, 1973-2003, *M≥5.0, n=47155; (b)* Japan, 1956-94, *M≥3.0, n=57047; (c)* California, SCSN catalog, 1975-2001, *M≥1.0, n=335641; (d)* Italy, 1956-97, *M≥2.5, n=18502; (e)* Greece, 1956-99, *M≥3.4, n=29824;* all reported with discrete magnitudes.

Fig. 13: *TED*-statistic as a function of the lower threshold $u$ for the distributions shown in Fig. 12. *(a)-(e)* correspond to Fig. 12.

Fig. 14: TED-statistic applied to the Harvard seismic moments (for all 14 subduction zones, n=4609) converted into discrete magnitudes $M_W$ in units of 0.1.

Fig. 15a: Tail distribution function of event sizes reported in the Harvard catalog in terms of magnitude $m_b$ determined from body wave measurements.

Fig. 15b: Tail distribution function of event sizes reported in the Harvard catalog in terms of magnitude $M = max(m_b ; M_s )$; $m_b$ is body wave magnitude; $M_s$ is surface wave magnitude.

Fig. 16a: *TED*-statistics for the magnitude distributions shown in Fig. 15a.

Fig. 16b: *TED*-statistics for the magnitude distributions shown in Fig. 15b.

*Table I. Characteristics of subduction zones; selected from (Jarrard,1986).*
*Values classified as "low" marked by (L), classified as "high" marked by (H).*

| Zone | Intermediate dip angle, *degrees* | Strain class, *(I – VII)* | Convergence rate, *cm/yr* | Slab age *m.y.* | Maximum observed magnitude, $M_w$ |
|---|---|---|---|---|---|
| Alaska | 18  (L) | VI  (H) | 6.3  (L) | 49  (L) | 9.1  (H) |
| Japan | 21  (L) | VI  (H) | 9.9  (H) | 67  (H) | 8.6  (H) |
| Kuril Isls | 28  (L) | V  (H) | 8.7  (H) | 119  (H) | 8.8  (H) |
| Kamchatka | 25  (L) | V  (H) | 8.8  (H) | 90  (H) | 9.0  (H) |
| Mariana | 26  (L) | IV  (L) | 7.6  (L) | 94  (H) | 7.2  (L) |
| Mexico | 60  (H) | VI  (H) | 7.2  (L) | 17  (L) | 8.4  (L) |
| S. America | 20  (L) | VII  (H) | 10.0  (H) | 38  (L) | 9.5  (H) |
| Sandwich Isls | 67  (H) | I  (L) | 0.9  (L) | 49  (L) | 7.0  (L) |
| New Hebrides | 44  (H) | I  (L) | 8.8  (H) | 52  (L) | 7.9  (L) |
| Solomon Isls | 42  (H) | IV  (L) | 12.0  (H) | 50  (L) | - |
| New Guinea | 35  (H) | I  (L) | 4.3  (L) | 50  (L) | - |
| Taiwan | 41  (H) | II  (L) | 3.0  (L) | 49  (L) | 8.0  (L) |
| Tonga | 29  (L) | I  (L) | 7.5  (L) | 117  (H) | 8.3  (L) |
| Sunda | 21  (L) | IV  (H) | 8.2  (H) | 88  (H) | 7.9  (L) |

**Fig.1**

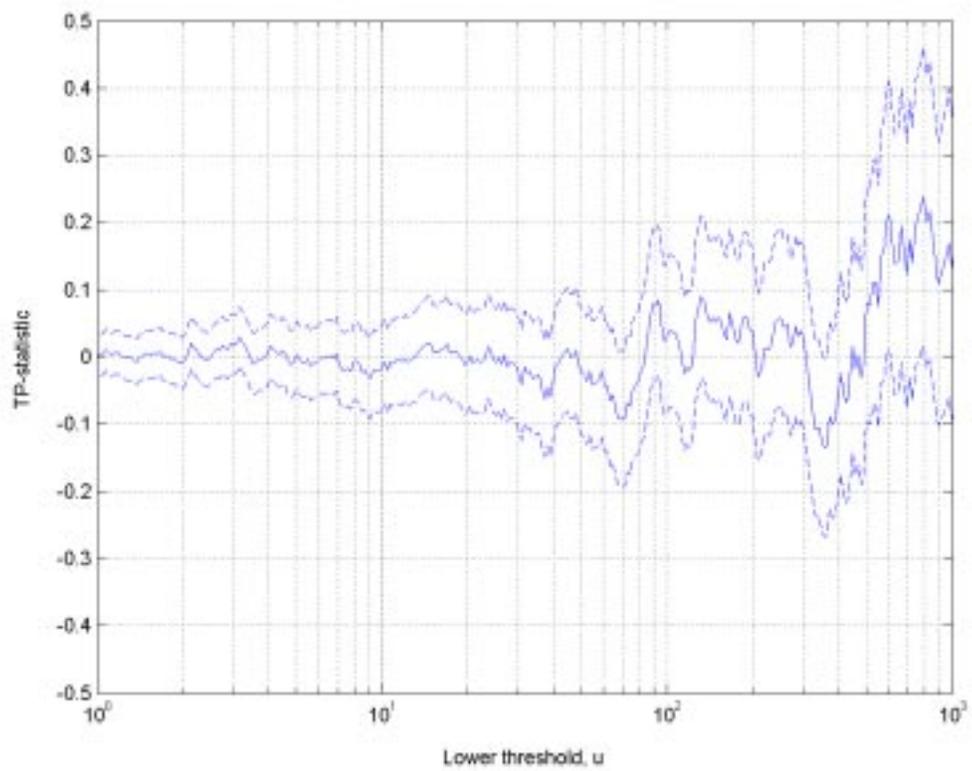

**Fig.2**

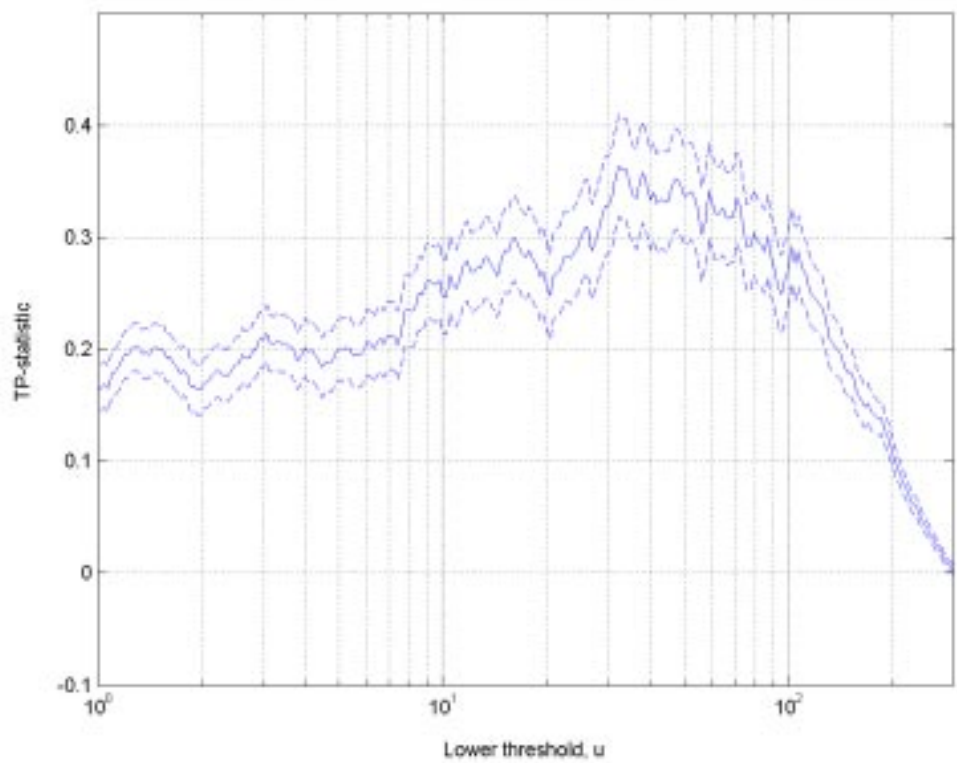

**Fig.3a**

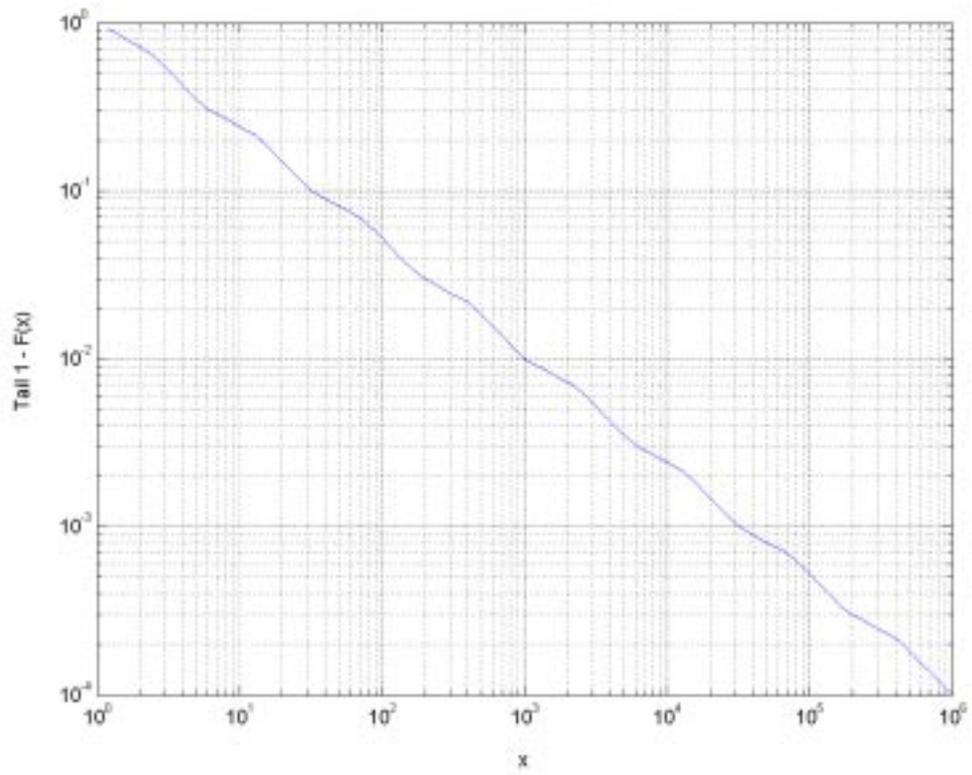

**Fig.3b**

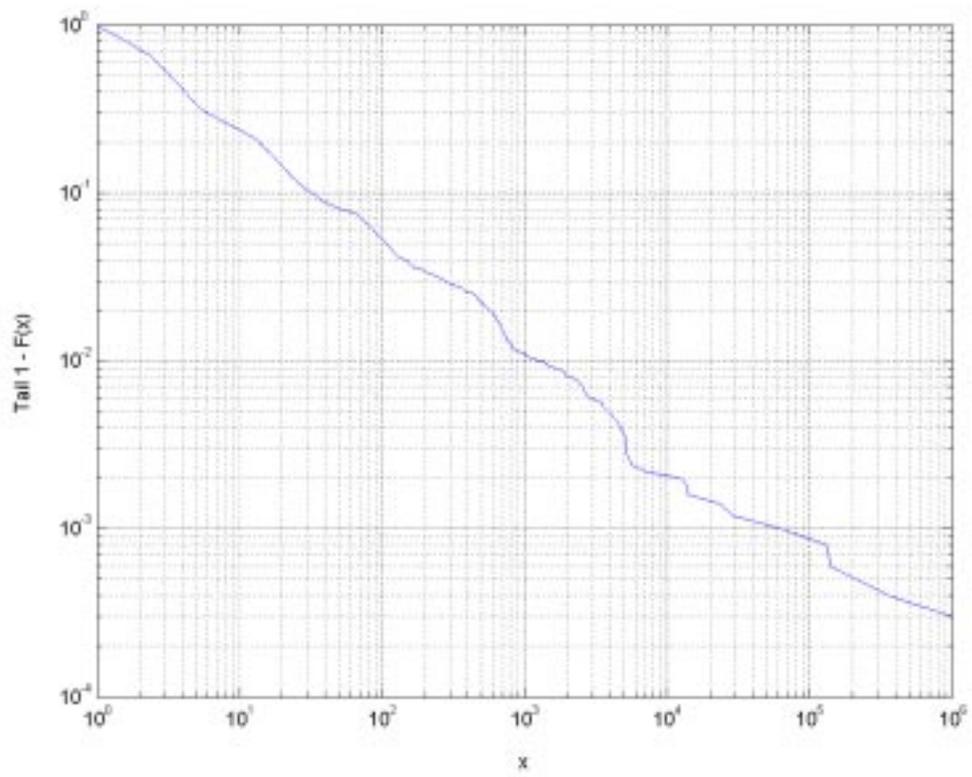

**Fig.3c**

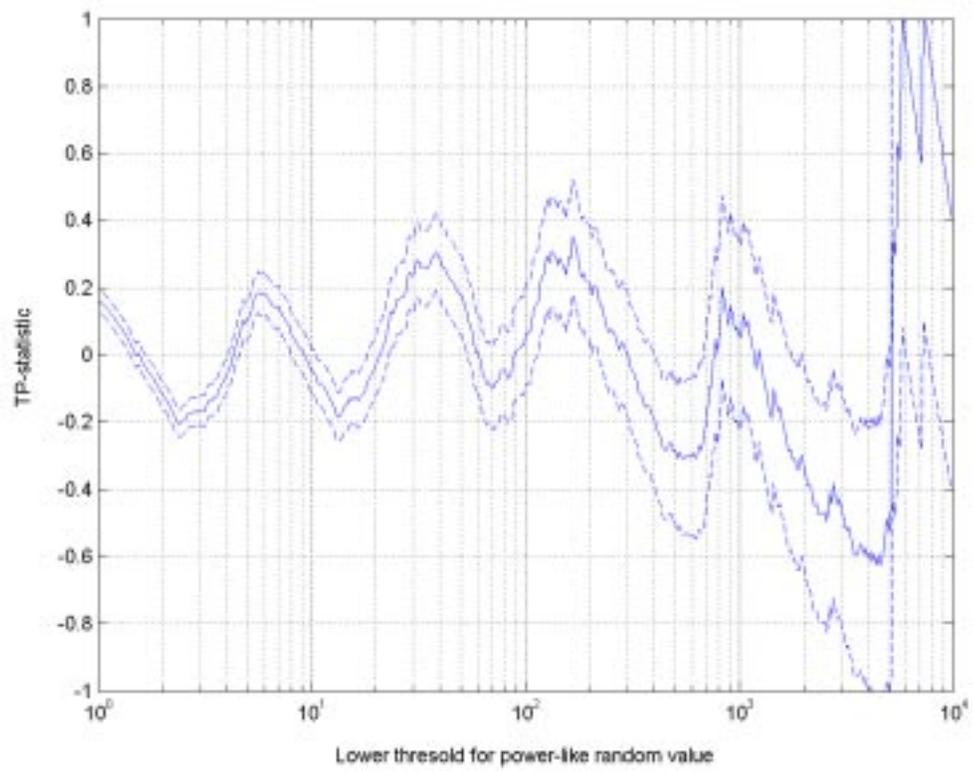

**Fig.4a**

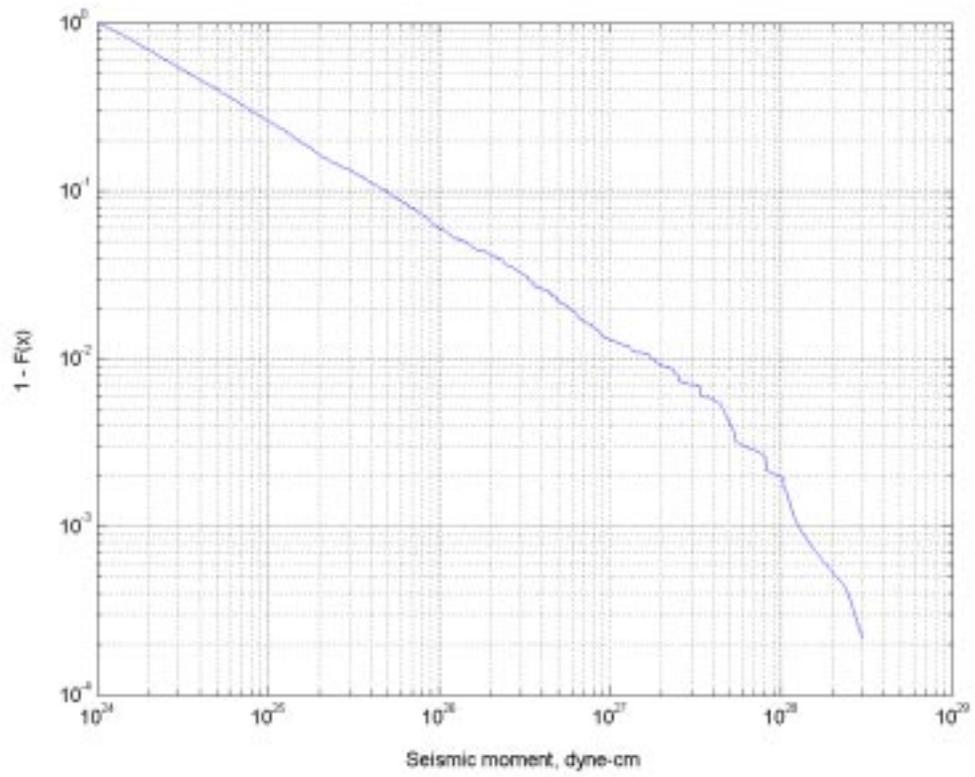

**Fig.4b**

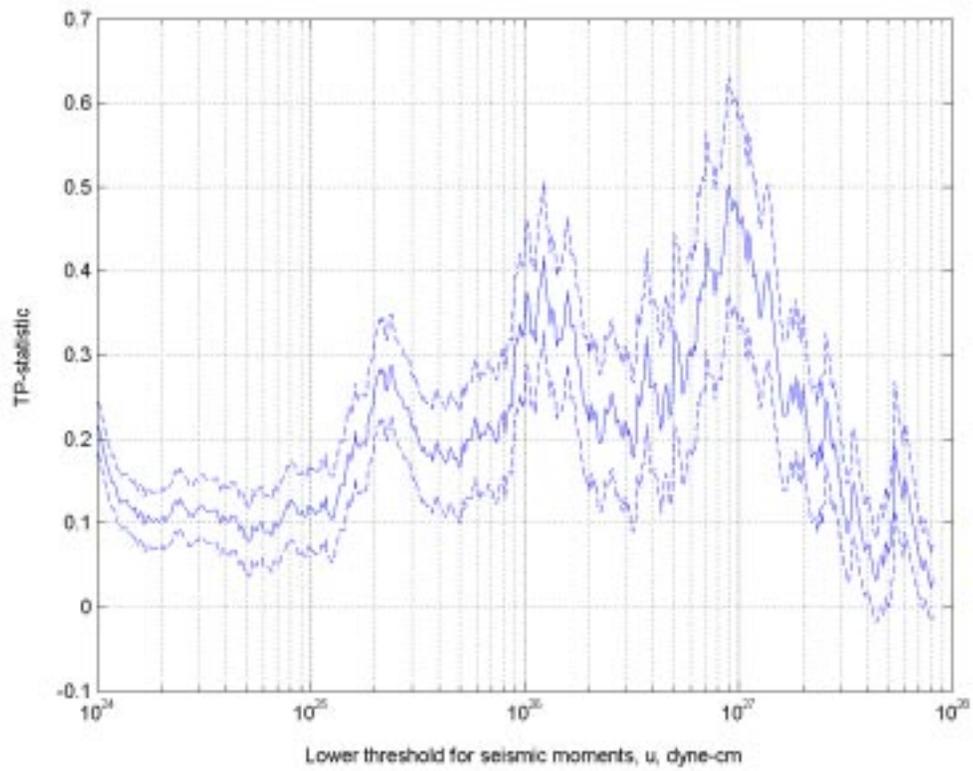

**Fig.4c**

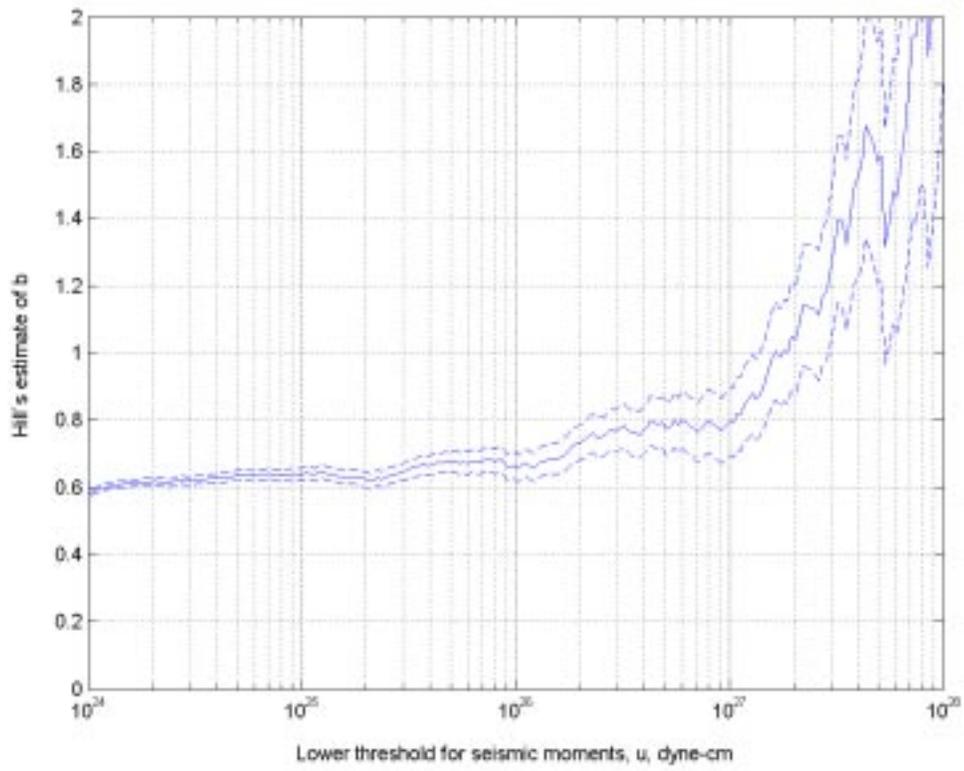

**Fig.5**

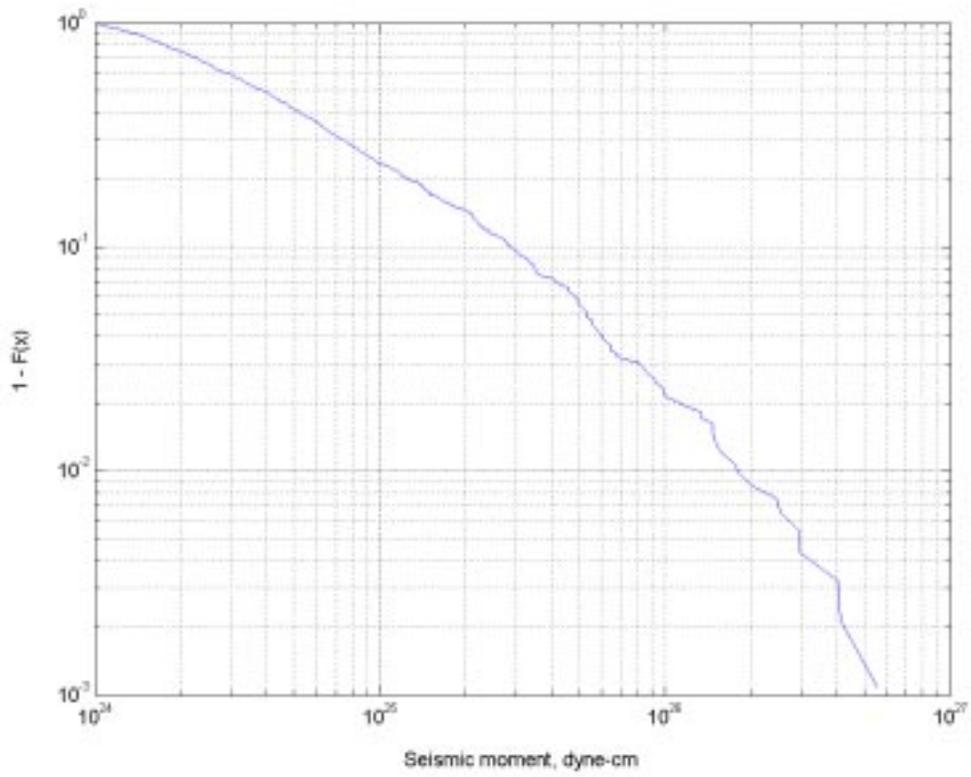

**Fig.6**

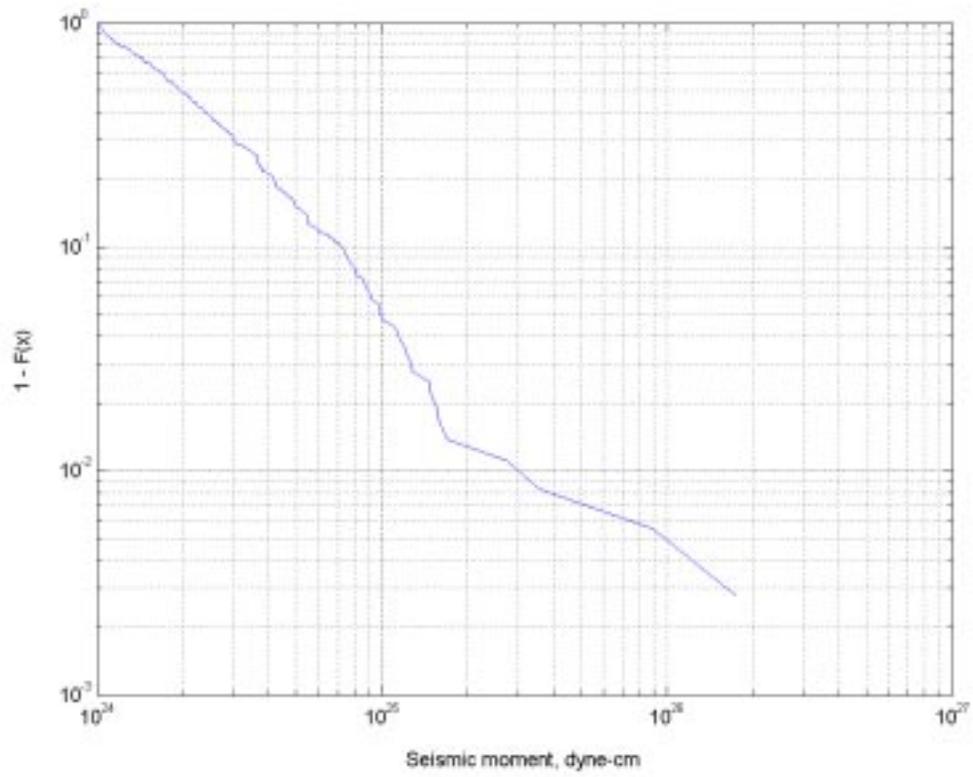

**Fig.7**

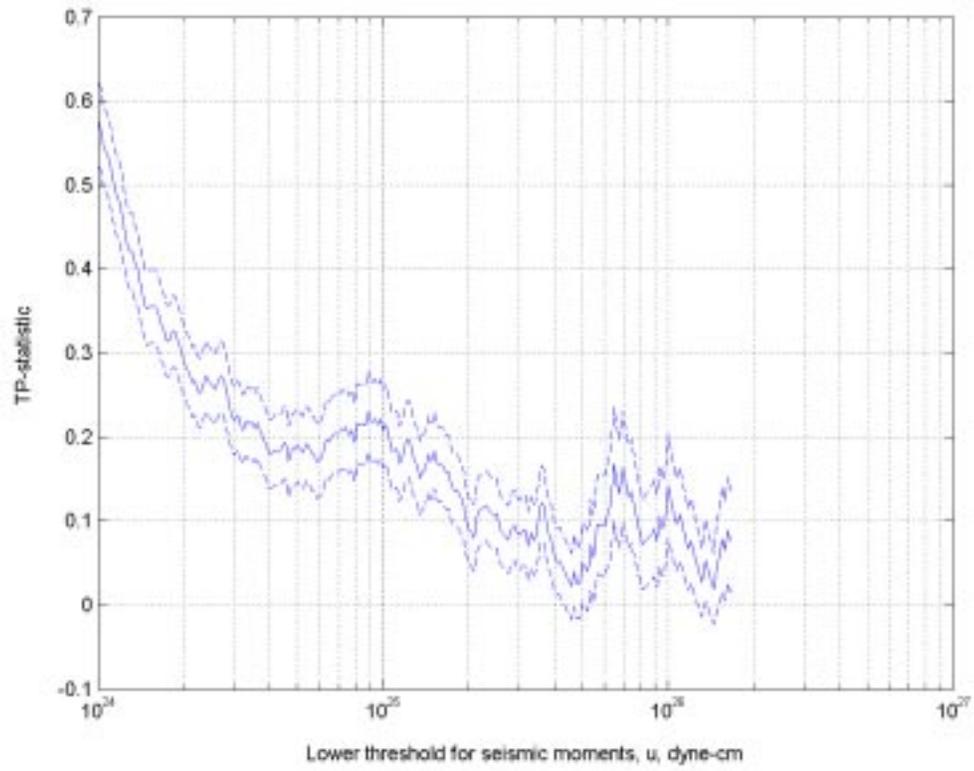

**Fig.8**

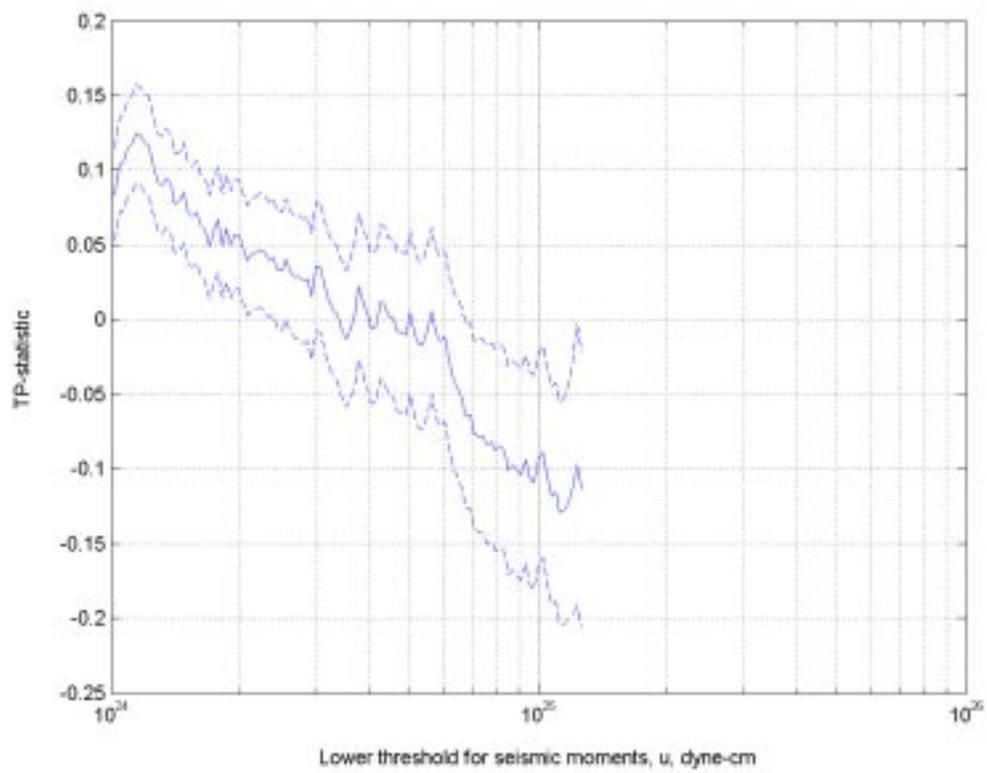

**Fig.9a**

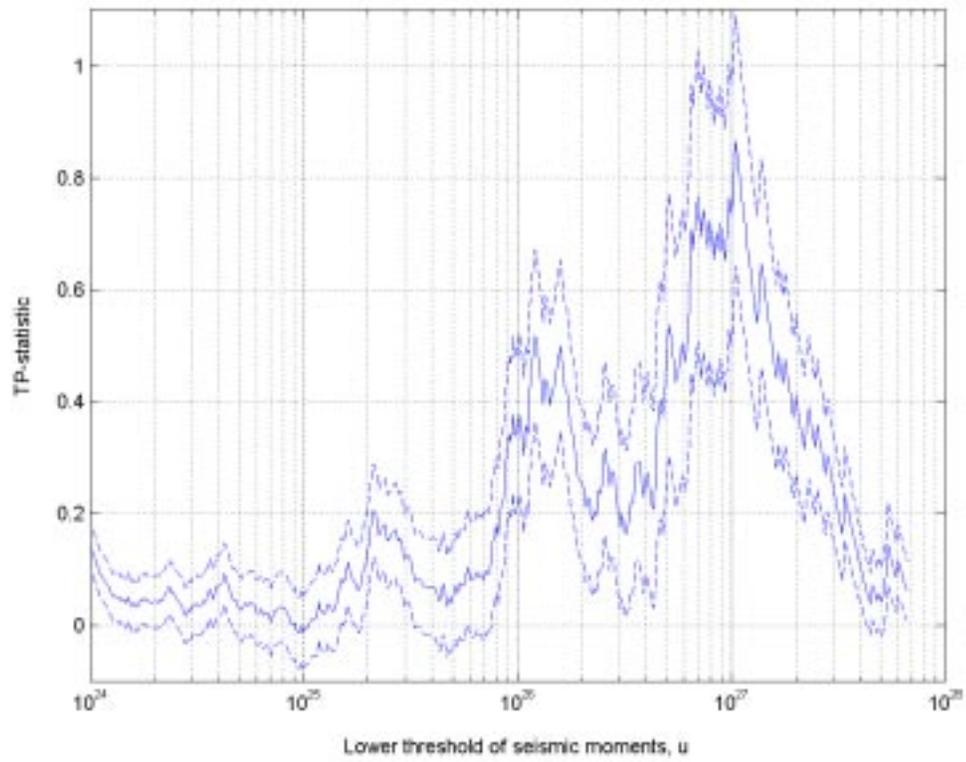

**Fig.9b**

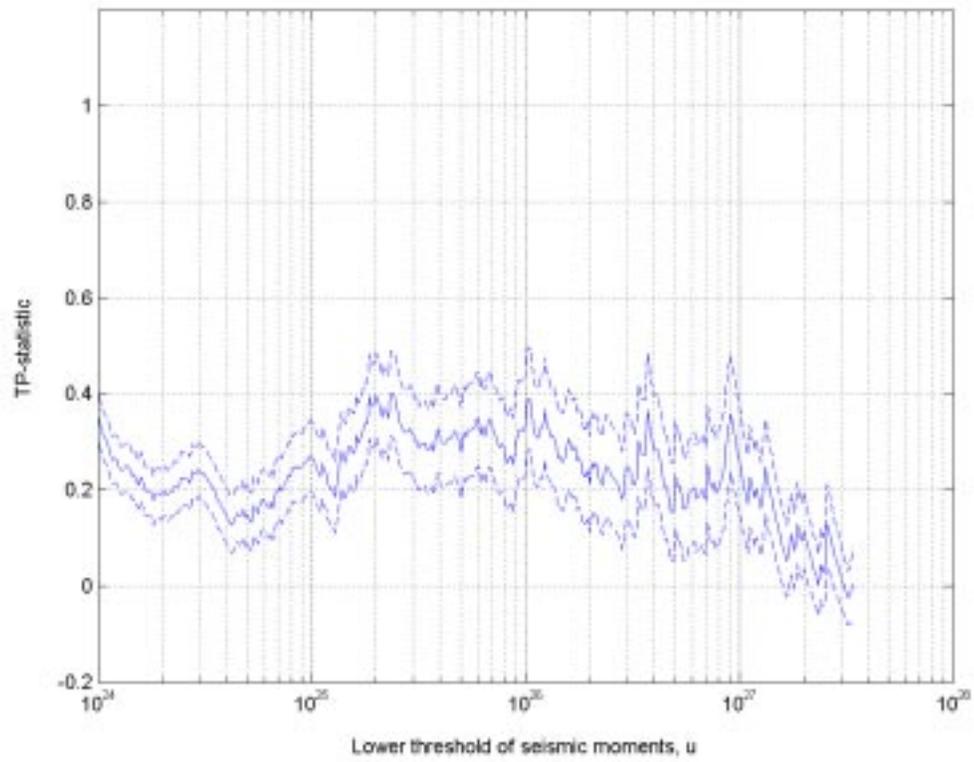

**Fig.10a**

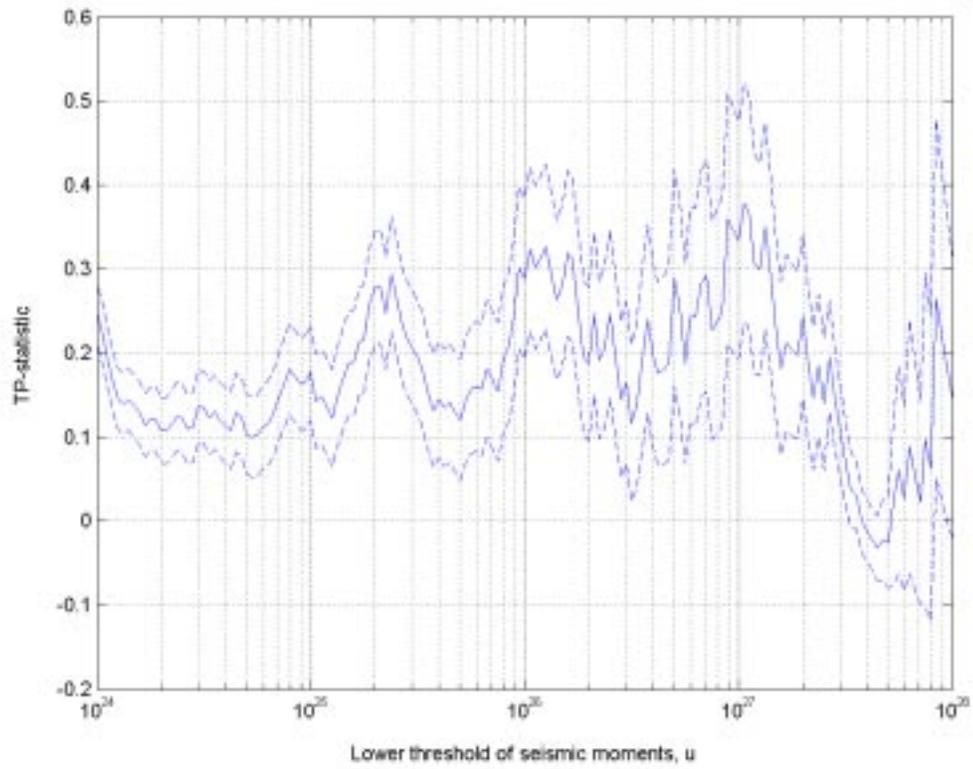

**Fig.10b**

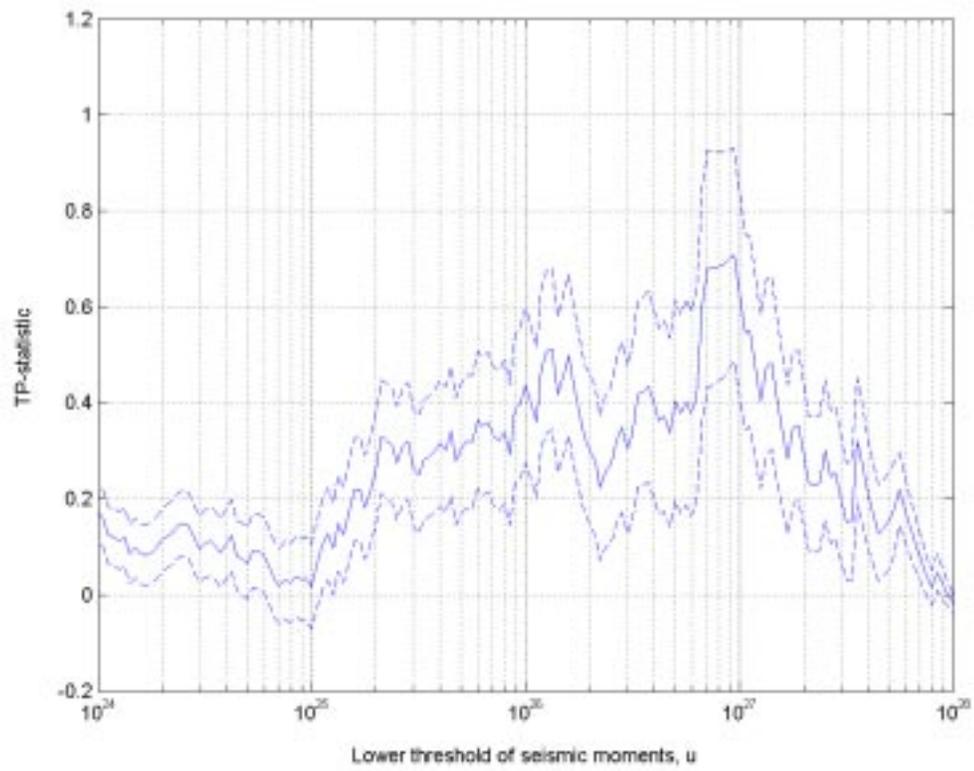

**Fig.11a**

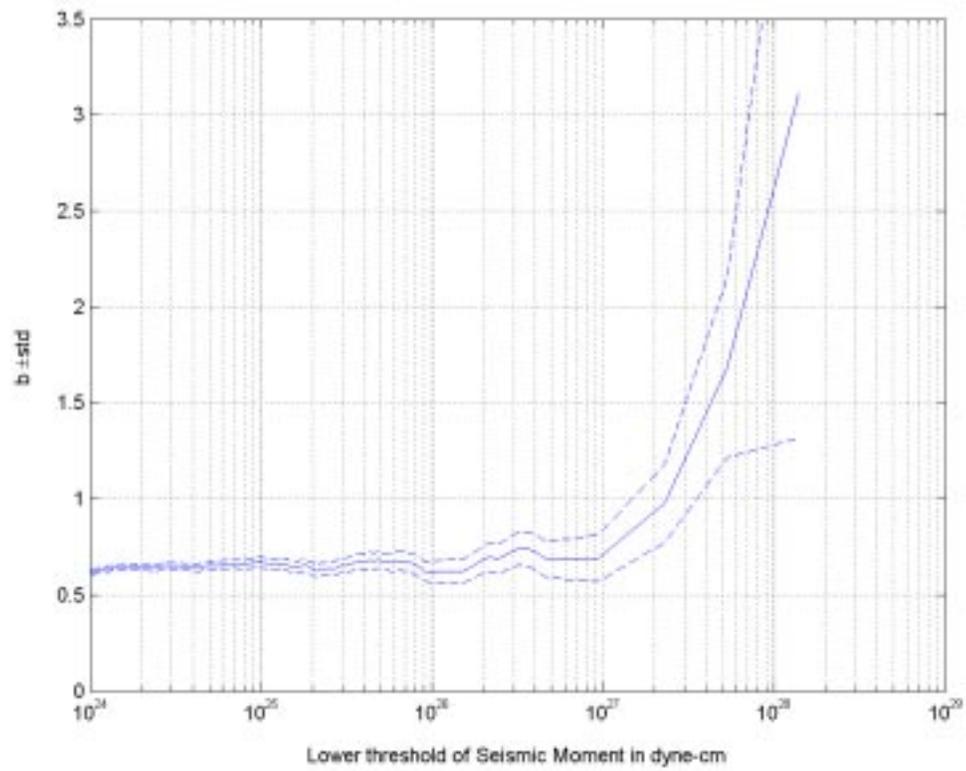

**Fig.11b**

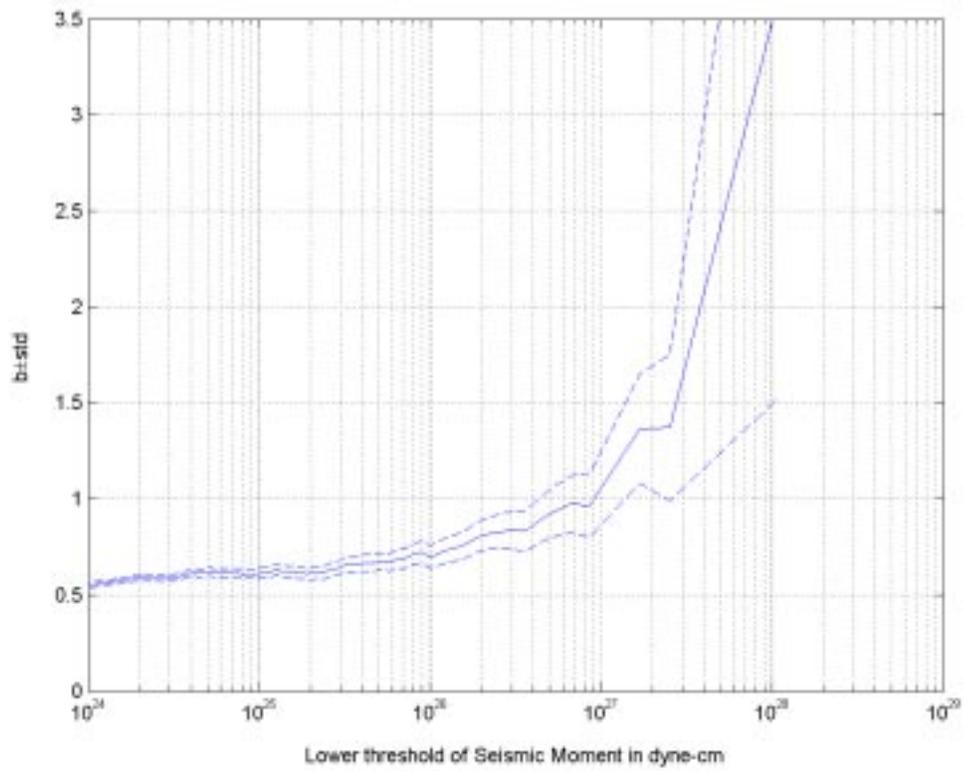

**Fig.12a**

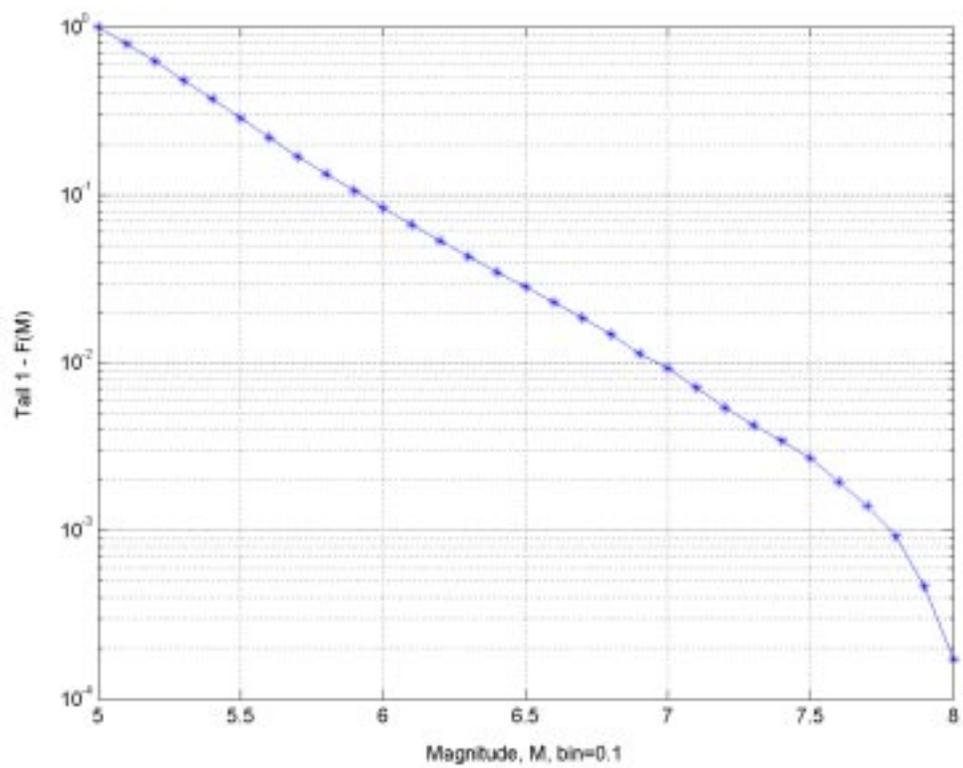

**Fig.12b**

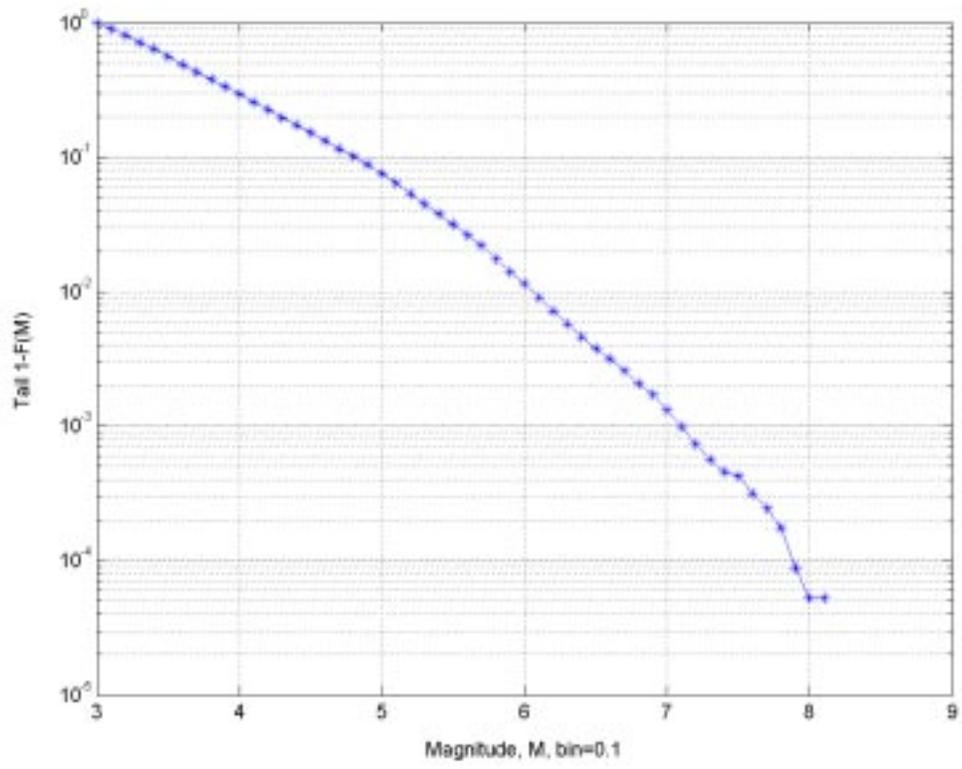

**Fig.12c**

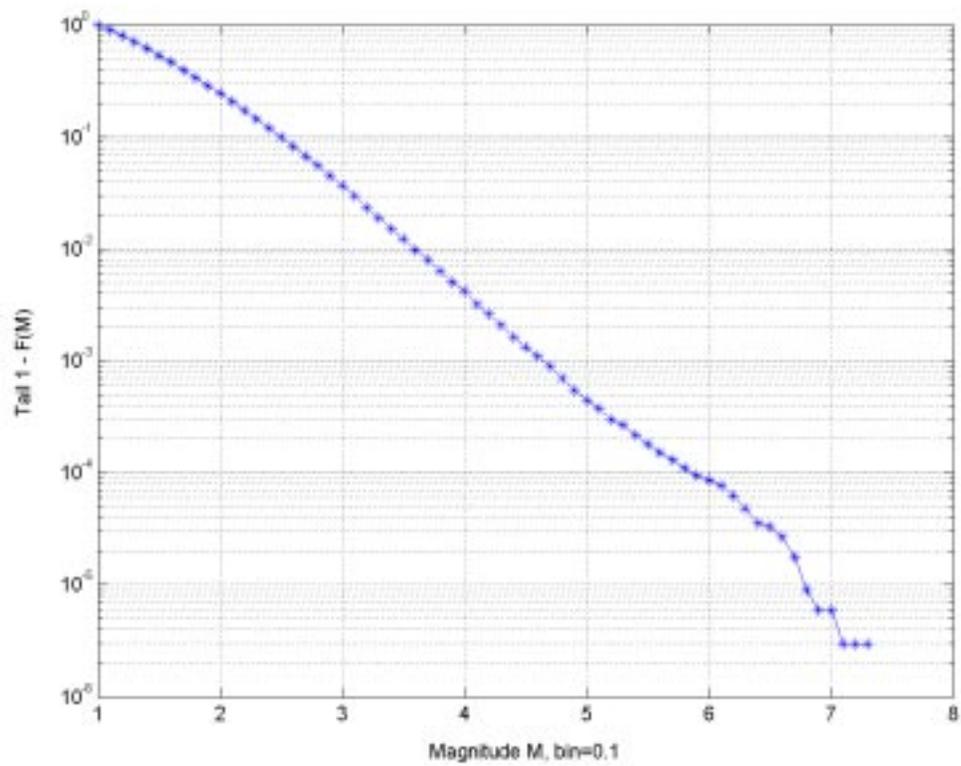

**Fig.12d**

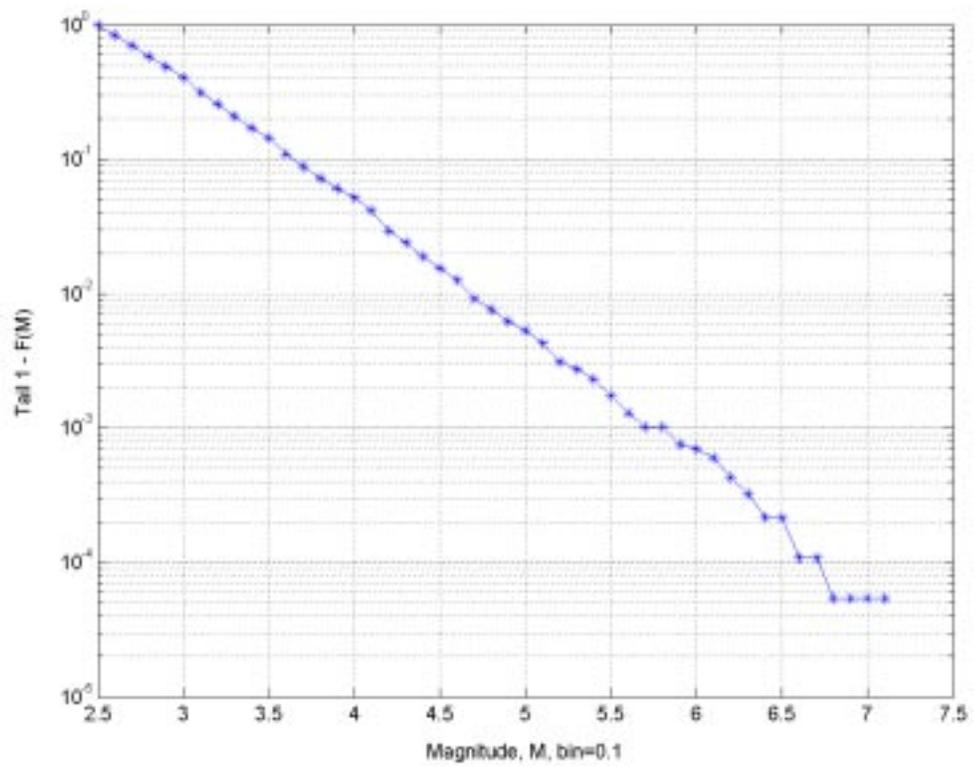

**Fig.12e**

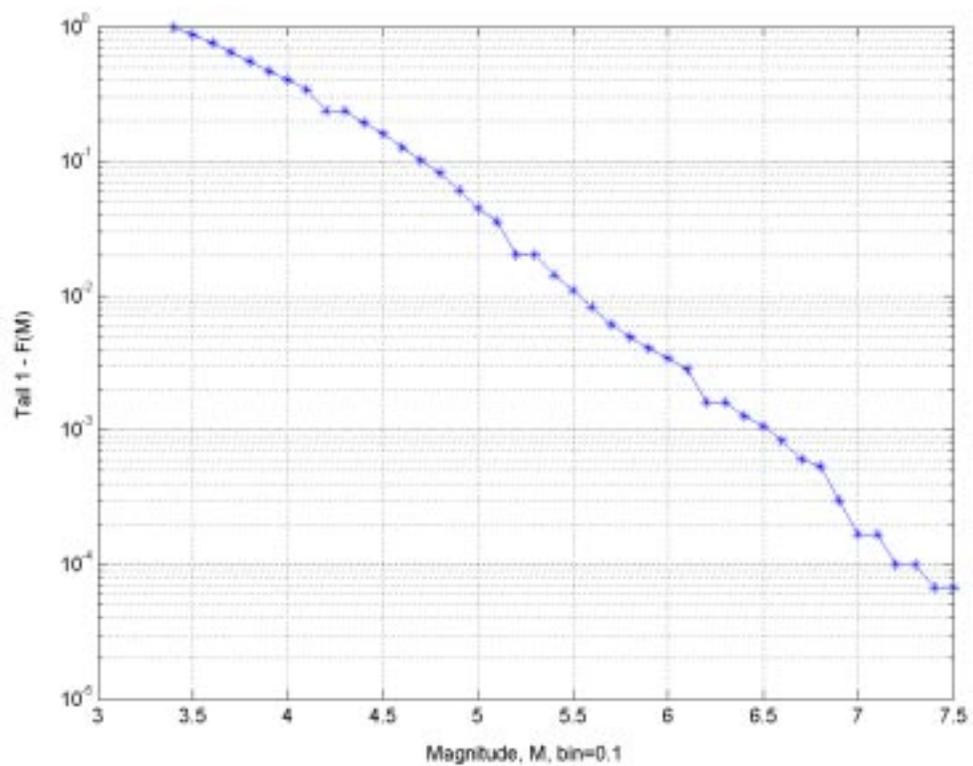

**Fig.13a**

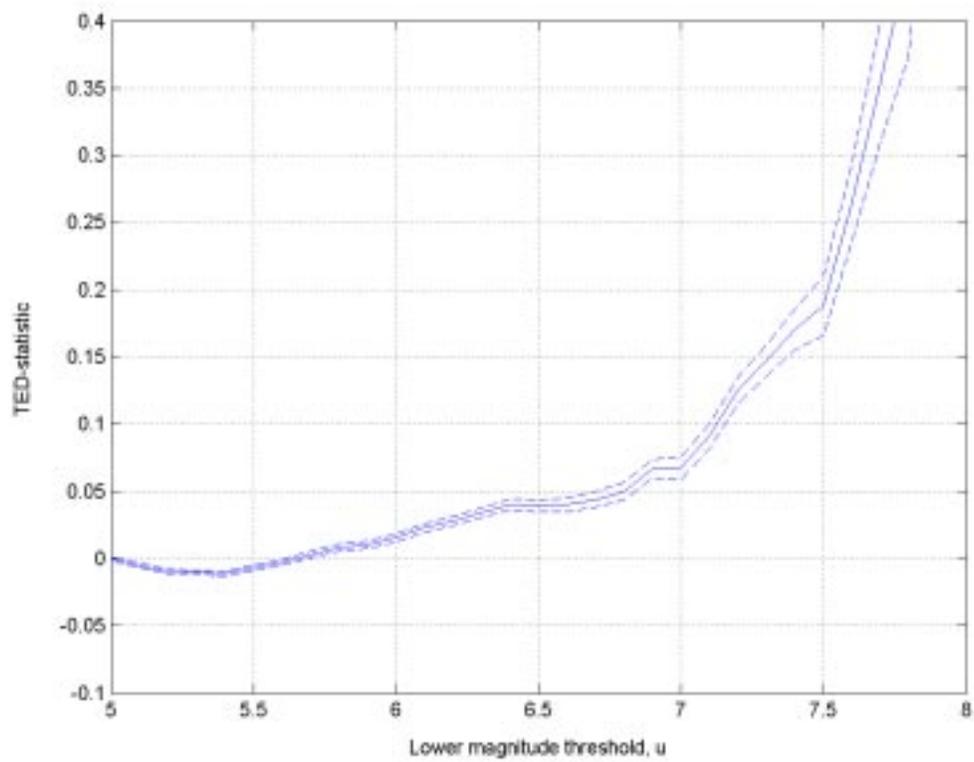

**Fig.13b**

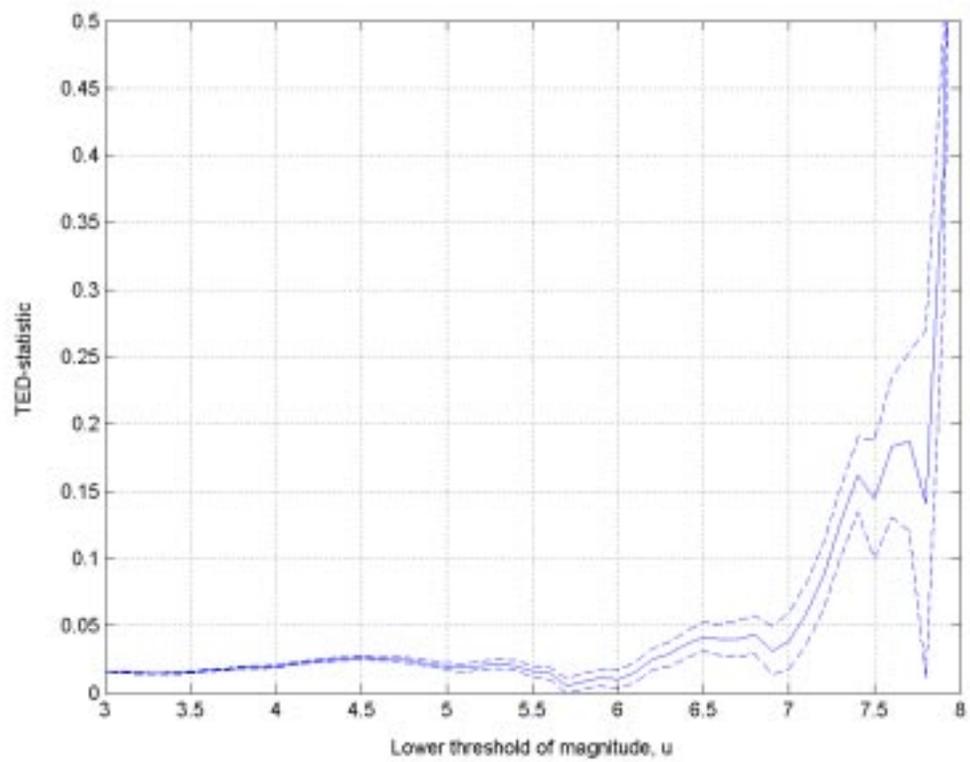

**Fig.13c**

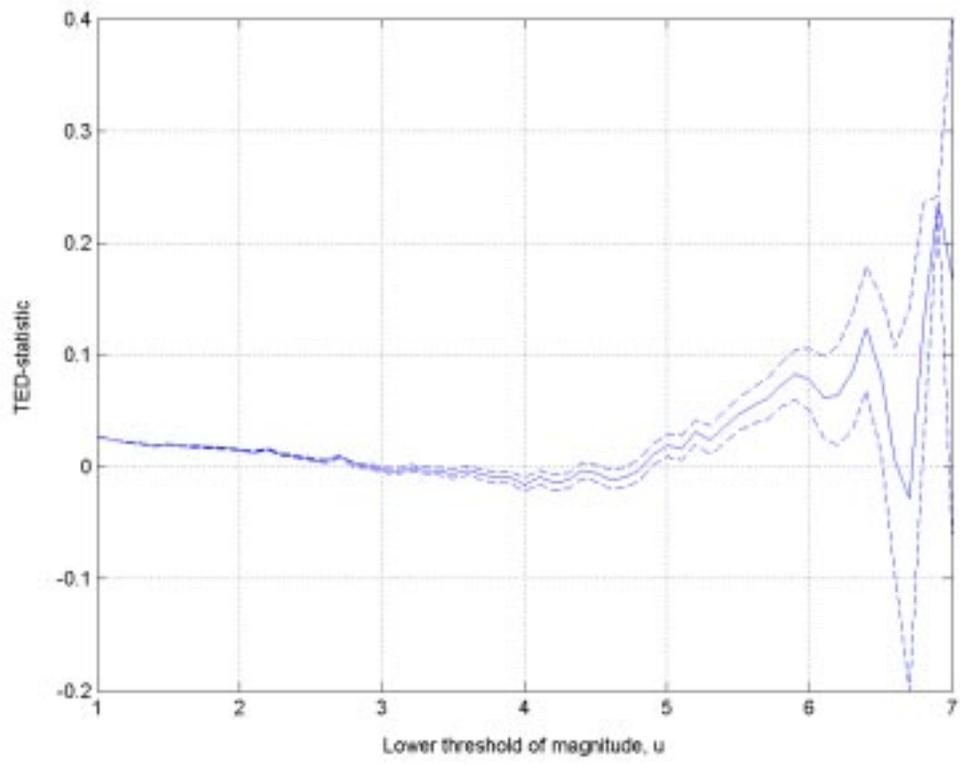

**Fig.13d**

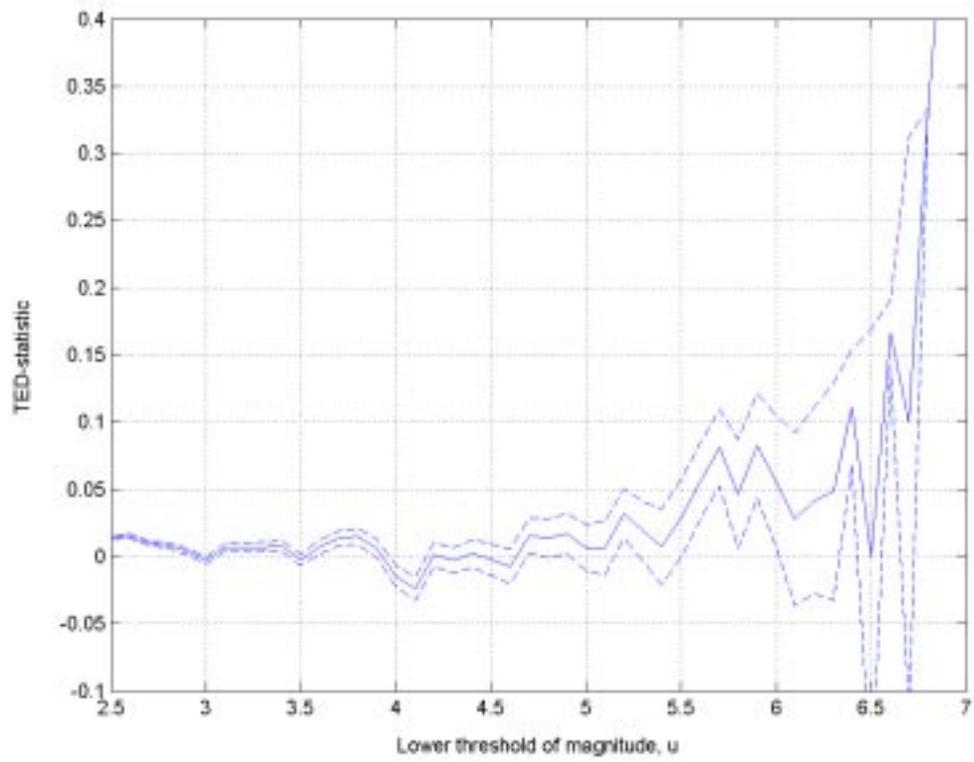

**Fig.13e**

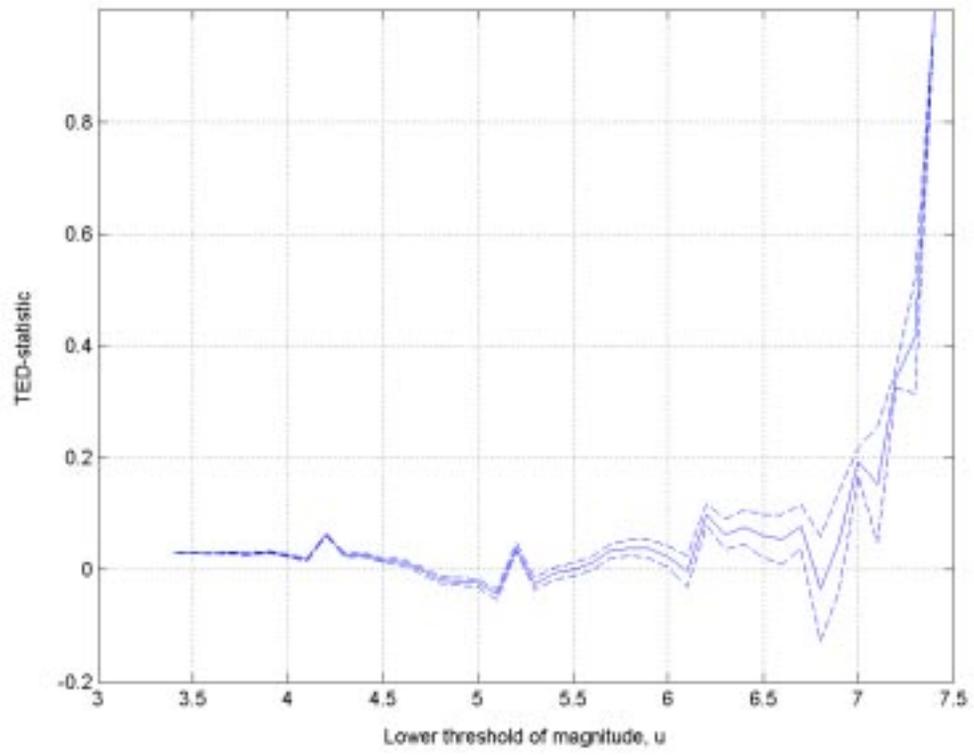

**Fig.14**

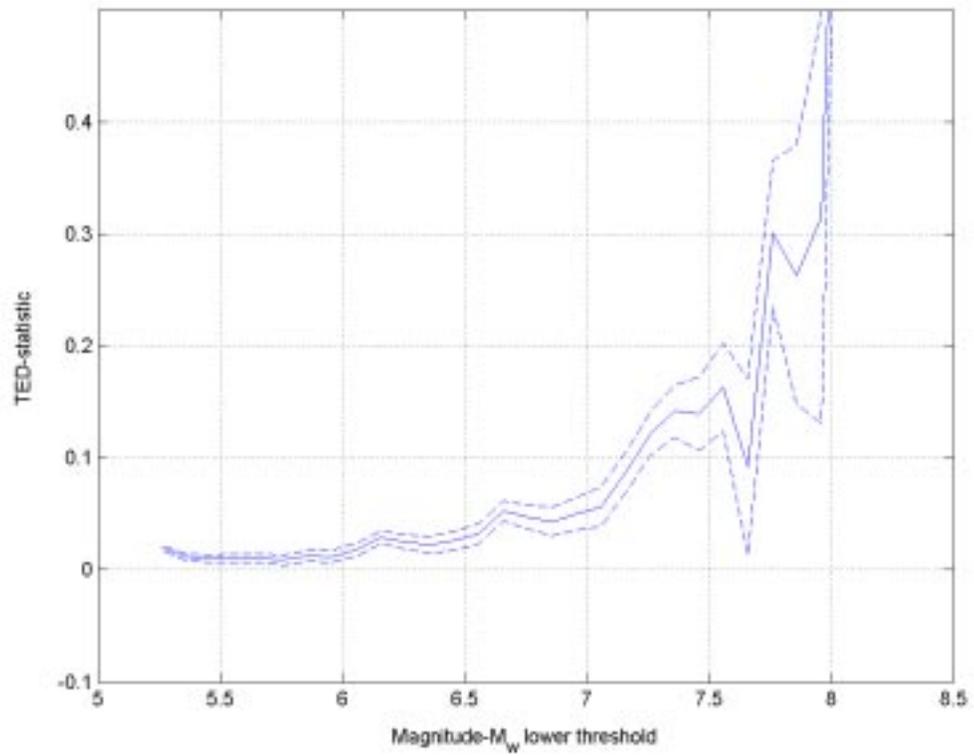

**Fig.15a**

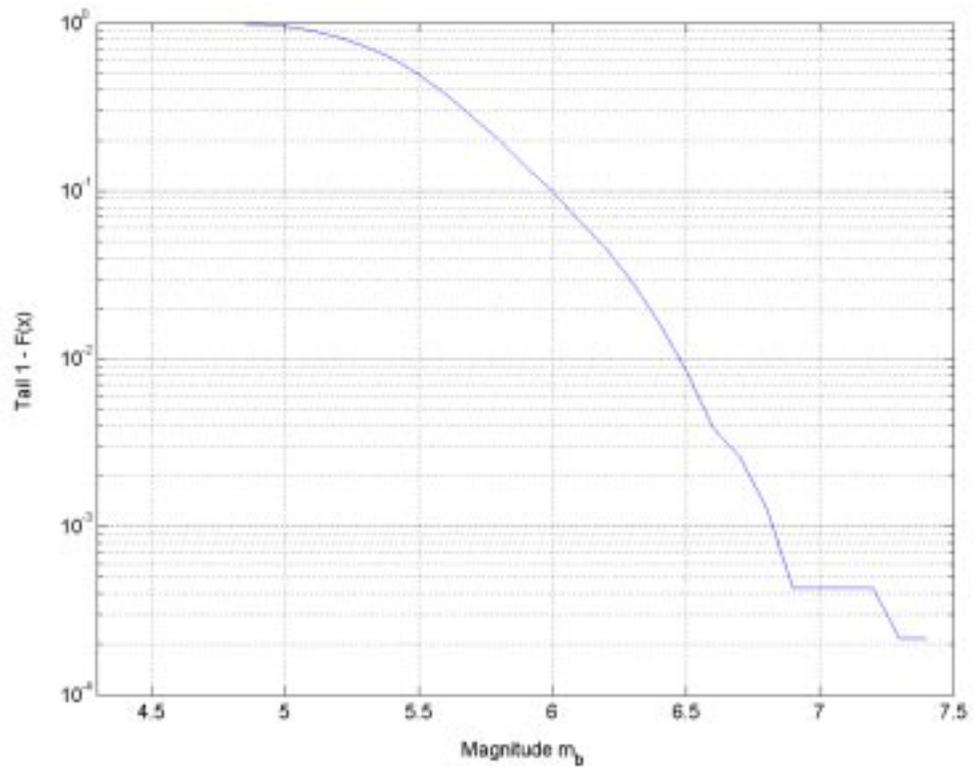

**Fig.15b**

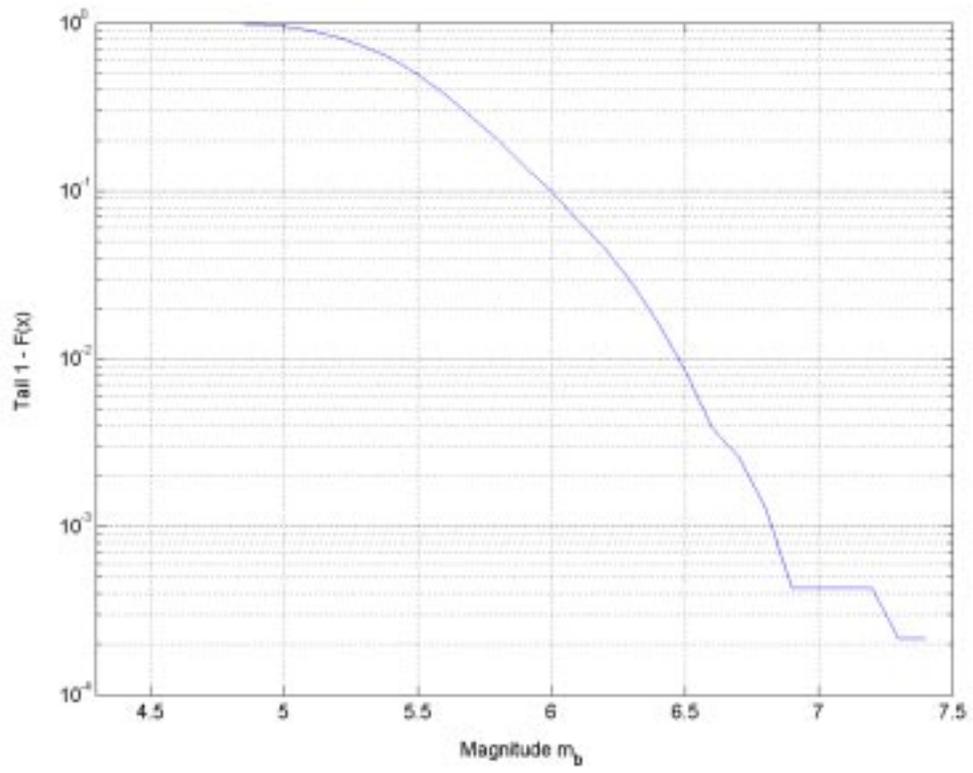

**Fig.16a**

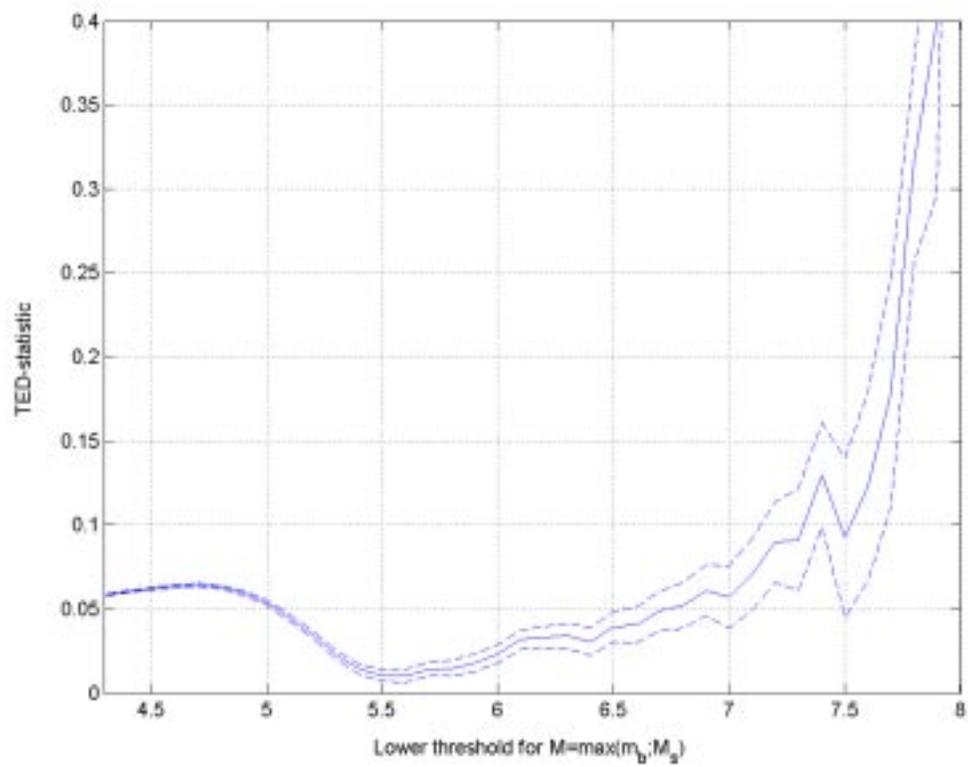

**Fig.16b**

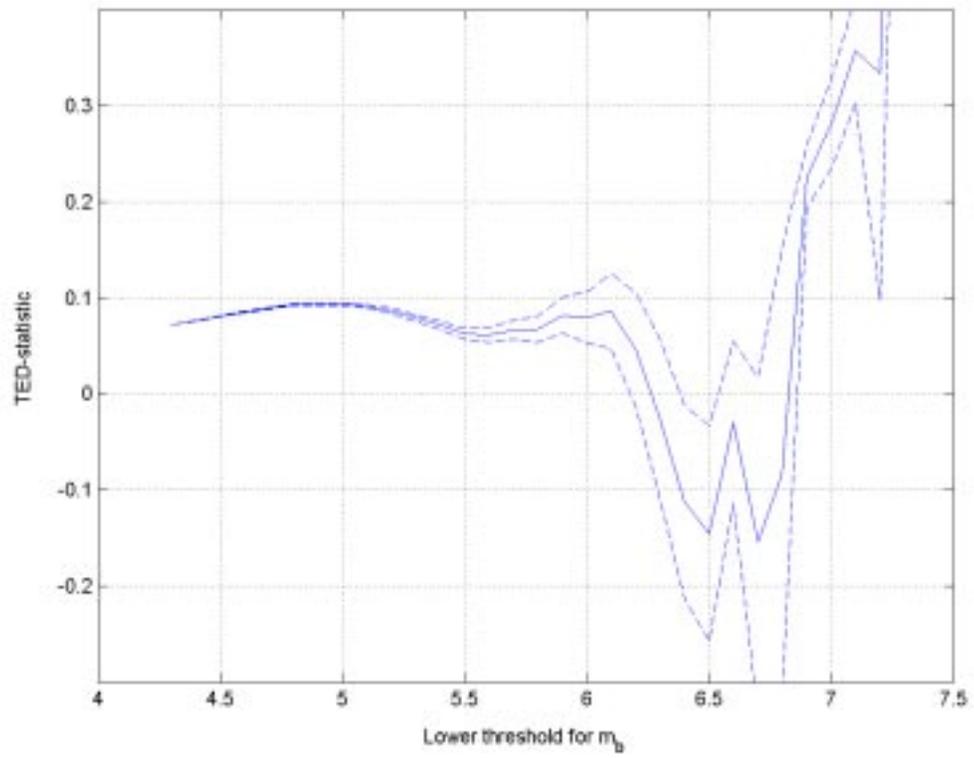